\documentclass[aps,amsmath,amssymb,twocolumn,showpacs,byrevtex,superscriptaddress]{revtex4}
\usepackage{graphicx}

\begin{document}

\title{The Role of Intramolecular Barriers on the Glass Transition of Polymers: \\
Computer Simulations vs. Mode Coupling Theory}

%
%

\author{Marco Bernabei}
\email[Corresponding author: ]{sckbernm@ehu.es}
\affiliation{Donostia International Physics Center, Paseo Manuel de Lardizabal 4,
20018 San Sebasti\'{a}n, Spain.}

\author{Angel J. Moreno}
\affiliation{Centro de F\'{\i}sica de Materiales (CSIC, UPV/EHU)-Materials Physics Center, 
Apartado 1072, 20080 San Sebasti\'{a}n, Spain.}

\author{Juan Colmenero}
\affiliation{Donostia International Physics Center, Paseo Manuel de Lardizabal 4,
20018 San Sebasti\'{a}n, Spain.}
\affiliation{Centro de F\'{\i}sica de Materiales (CSIC, UPV/EHU)-Materials Physics Center, 
Apartado 1072, 20080 San Sebasti\'{a}n, Spain.}
\affiliation{Departamento de F\'{\i}sica de Materiales, Universidad del Pa\'{\i}s Vasco (UPV/EHU),
Apdo. 1072, 20080 San Sebasti\'{a}n, Spain.}

\begin{abstract}

We present computer simulations of a simple bead-spring model for polymer melts with 
intramolecular barriers. By systematically tuning the strength of the barriers, 
we investigate their role on the glass transition. Dynamic observables
are analyzed within the framework of the Mode Coupling Theory (MCT). Critical nonergodicity parameters, critical
temperatures and dynamic exponents are obtained from consistent fits of simulation data to MCT asymptotic laws. 
The so-obtained MCT $\lambda$-exponent increases from standard values for fully-flexible chains
to values close to the upper limit for stiff chains. In analogy with systems exhibiting higher-order MCT transitions,
we suggest that the observed large $\lambda$-values arise form the interplay between two distinct mechanisms
for dynamic arrest: general packing effects and polymer-specific intramolecular barriers. 
We compare simulation results with numerical solutions of the MCT equations for polymer systems, within
the polymer reference interaction site model (PRISM) for static correlations. We verify
that the approximations introduced by the PRISM are fulfilled by simulations, 
with the same quality for all the range of investigated barrier strength.
The numerical solutions reproduce the qualitative trends of simulations for the dependence
of the nonergodicity parameters and critical temperatures on the barrier strength. In particular,
the increase of the barrier strength at fixed density increases the localization length
and the critical temperature. However the qualitative agreement between theory and simulation
breaks in the limit of stiff chains. We discuss the possible origin of this feature.

\end{abstract}
\date{\today}
\maketitle

\begin{center}
{\bf I. INTRODUCTION}
\end{center}

Since they do not easily crystallize, polymers are probably the most extensively 
studied systems in relation with the glass transition phenomenon.
Having said this, their macromolecular character, and in particular chain connectivity,
must not be forgotten. The most evident effect of chain connectivity is the sublinear increase 
of the mean squared displacement (Rouse-like) \cite{doi}
arising after the decaging process, in contrast to 
the linear regime found in non-polymeric glass-formers.
Moreover, in the case of strongly entangled polymer chains, the reptation
model predicts other two sublinear regimes between
the Rouse and linear regimes \cite{doi,degennes,mcleish}.

Another particular ingredient of polymers is that, apart from fast librations 
or methyl group rotations \cite{methylrev}, 
every motion involves jumps over carbon-carbon rotational barriers and/or chain conformational changes. 
Intramolecular barriers play a decisive role in the physical properties of polymer systems.
Thus, they are responsible of partial or total crystallization \cite{meyer,vettorel}.
They also enhance dynamic features which are usually associated to reptation
\cite{faller,bulacu1}, which controls rheological properties \cite{mcleish}.
Models for semiflexible polymers are of great interest, since they can be applied to many important biopolymers 
such as proteins, DNA, rodlike viruses, or actin filaments
\cite{bustamante,kas,ober}.
Moreover, chain stiffness seems to play an important role in the absorption
behavior of polymers at interfaces \cite{semenov,ubbink}. 
Thus, an understandig of the role of intramolecular barriers on structural, dynamic
and rheological properties of polymers is of practical as well as of fundamental interest.

In this work we investigate, by means of molecular dynamics simulations, 
the role of intramolecular barriers on the glass transition
of polymer melts, by systematically tuning barrier strength in a simple bead-spring model.
We discuss the obtained results within the framework of the Mode Coupling Theory (MCT) 
of the glass transition \cite{mctrev1,mctrev2,gotzebook,reichman,das,gotzenewbook}. 
We extend preliminary results reported 
by us in Ref. \cite{bernabei} by testing a large set of predictions, including the factorization theorem
and time-temperature superposition principle. A consistent set of dynamic
exponents associated to asymptotic scaling laws is obtained.
By increasing the barrier strength a crossover is observed for the values of the so-called
$\lambda$-exponent. In the limit of fully-flexible chains $\lambda$ takes values $\sim 0.7$,
characteristic of simple fluids dominated by packing effects. On the contrary,
for strong intramolecular barriers the $\lambda$-values approach the upper limit $\lambda =1$
characteristic of higher-order MCT transitions. The latter arise in systems with
different coexisting mechanisms for dynamic arrest \cite{schematic,sperl,krakoviack}. 
In the system investigated here, the obtained results suggest an interplay between 
general packing effects and polymer-specific intramolecular barriers.

Chong and co-workers \cite{chongprl,chongpre} have recently presented an extension of the MCT to simple 
fully-flexible bead-spring models of polymer systems, in the framework
of the polymer reference interaction site model (PRISM) \cite{schweizer,curro,putz}. 
In this formalism each molecule is divided into interaction sites corresponding to monomers. 
A key assumption of the PRISM is
the replacement of the site-specific intermolecular surroundings of a monomer by an averaged one 
(equivalent site approximation), while keeping the fully intramolecular dependence.
We have tested the PRISM approximations used by MCT in the polymer model here investigated, 
which incorporates intramolecular barriers. Likewise, we have solved the MCT equations
for the location of the MCT `glass transition' temperatures (MCT critical temperatures) 
and for the nonergodicity parameters, which quantify
the stability of density fluctuations in the reciprocal space. 
We compare solutions of the MCT equations with the results obtained from the phenomenological 
analysis of the simulation data. We observe that the theory reproduces  qualitative  
trends in the nonergodicity parameters and critical temperatures. However, the agreement breaks 
as the limit of stiff chains is approached. We discuss the possible origins of this feature.  

The article is organized as follows. In Section II we describe the model and give simulation details.
Static correlators are shown in Section III. Moreover the PRISM approximations are tested 
for representative values of the barrier strength. Section IV presents qualitative dynamic trends
as a function of the barrier strength. In Section V we summarize the universal predictions of the MCT 
and the equations of motion of the version for polymer melts introduced by Chong and co-workers.
In Section VI we perform a phenomenological analysis of simulation data within the MCT, by testing
universal scaling laws and deriving their associated dynamic exponents. In Section VII we compare
the results of the former analysis with numerical solutions of the MCT equations.
We discuss the observed differences for stiff chains in Section VIII. 
Conclusions are given in Section IX.

\begin{center}
{\bf II. MODEL AND SIMULATION DETAILS}
\end{center}

We have performed molecular dynamics (MD) simulations of a bead-spring  model 
for which we have implemented bending and torsional intramolecular barriers.
The monomer-monomer interaction is given  by a corrected soft-sphere potential
\begin{equation}
V(r) = 4\epsilon[(\sigma/r)^{12} - C_0 + C_2(r/\sigma)^{2}],
\label{eq:potsoft}
\end{equation}
where $\epsilon=1$ and $\sigma=1$. The potential $V(r)$ is set to zero beyond 
the cutoff distance $r \geq c\sigma$, with $c = 1.15$.
The values $C_0 = 7c^{-12}$ and $C_2 = 6c^{-14}$  guarantee continuity of  potential and forces at $r=c\sigma$. 
The potential $V(r)$ is purely repulsive. It does not show local minima within the interaction range $r<c\sigma$. 
Thus, it drives dynamic arrest only through packing effects. 
Along the chain backbone, of $N$ monomers, an additional finitely-extensible nonlinear elastic (FENE)  
potential \cite{grest,bennemann} is used to introduce bonds between consecutive monomers:
\begin{equation}
V_{\rm FENE}(r) = -\epsilon K_{\rm F}R_0^2 \ln[ 1-(R_0\sigma)^{-2}r^2 ],
\label{eq:potfene}
\end{equation}
where $K_F=15$ and $R_0=1.5$. The superposition of potentials (\ref{eq:potsoft}) and (\ref{eq:potfene}) 
provides an effective bond potential for consecutive monomers
with a sharp minimum at $r\approx 0.985$, which makes bond crossing impossible.

Intramolecular barriers are implemented by means of a combined bending $V_B$, and torsional potential $V_T$.
We have used the potentials  proposed by Bulacu and van der Giessen in Refs. \cite{bulacu1,bulacu2}.
The bending potential acts on three consecutive monomers along the chain. The angle between adjacent pairs 
of bonds is mantained close to the equilibrium value $\theta_0 = 109.5^{\rm o}$ by the cosine harmonic bending potential
\begin{equation}
V_{\rm B}(\theta_i) = (\epsilon K_{\rm B}/2 )(\cos\theta_i - \cos\theta_0)^2,
\label{eq:potben}
\end{equation}
where $\theta_i$ is the bending angle between consecutive monomers $i-1$, $i$ and $i+1$ (with $2 \leq i \leq N-1$).

The torsional  potential constrains the dihedral angle $\phi_{i,i+1}$, which is defined for the consecutive monomers  
$i-1$, $i$, $i+1$ and $i+2$ (with $2 \leq i \leq N-2$), as the angle between the two planes defined by the sets
($i-1$, $i$, $i+1$) and ($i$, $i+1$, $i+2$). The form of this potential is
\begin{eqnarray}
V_{\rm T}(\theta_{i},\theta_{i+1},\phi_{i,i+1}) = \hspace{2 cm} \nonumber \\
\epsilon K_{\rm T}\sin^3 \theta_{i} \sin^3 \theta_{i+1} \sum_{n=0}^3 a_n \cos^n \phi_{i,i+1}.
\label{eq:pottor}
\end{eqnarray}
The values of the coefficients $a_n$ are $a_0=3.00$, $a_1=-5.90$, $a_2=2.06$, and $a_3=10.95$ \cite{bulacu1,bulacu2}.
The torsional potential depends both on the dihedral angle $\phi_{i,i+1}$  
and on the bending angles $\theta_i$ and $\theta_{i+1}$. As noted in Refs. \cite{bulacu1,bulacu2},
numerical instabilities arising when two consecutive bonds align are naturally eliminated
by choosing the torsional potential (\ref{eq:pottor}),
without the need of imposing rigid constraints on the bending angles.

In the following, temperature $T$, time $t$, distance, wave vector $q$, and monomer density $\rho$ 
are given respectively in units of $\epsilon/k_B$ (with $k_B$ the Boltzmann constant), 
$\sigma(m/\epsilon)^{1/2}$ (with $m$ the monomer mass), $\sigma$, $\sigma^{-1}$, and $\sigma ^{-3}$.
We investigate, at fixed monomer density $\rho = 1.0$, the  temperature dependence of the dynamics 
for different values of the bending and torsion strength,
$(K_{\rm B}$,$K_{\rm T}) =$ (0,0), (4,0.1), (8,0.2), (15,0.5), (25,1), (25,4), and (35,4).
In the following, all the data presented in the figures and discussed in the main text will correspond
to $\rho = 1.0$. This value will not be, in general, explicitly mentioned there. We have also
studied the case $(K_{\rm B}$,$K_{\rm T}) =(35,4)$ at density $\rho = 0.93$. The specific 
information of this case is given in Table \ref{tab:paramkbkt} (see below).
We investigate typically 8-10 different temperatures for each set of values  $(K_{\rm B}$,$K_{\rm T})$.

We simulate 300 chains, each chain consisting of $N = 10$ monomers of mass $m=1$, 
placed in a cubic simulation box of lenght  $L_{\rm box}=14.4225$
for $\rho =1.0$, or $L_{\rm box}=14.7756$ for $\rho =0.93$, with periodic boundary conditions.
Equations of motion are integrated by using the velocity Verlet scheme \cite{frenkel}.
Computational expense is reduced by implementing a linked-cell method \cite{frenkel}.
We use a  time step ranging from $10^{-4}$ to $5 \times 10^{-3}$. We take shorter and longer steps 
for  respectively higher and lower values of temperatures, bending and torsional constants.
The system is prepared by placing and growing the chains randomly in the simulation box, 
with a constraint avoiding monomer core overlap. The initial monomer density is  $\rho=0.375$. 
Equilibration consists of a first run where the  box is rescaled
periodically by a factor $0.99< f < 1$ until the target density $\rho$ is reached, and 
a second isochoric run at that $\rho$. Thermalization at the target $T$ is achieved by 
periodic velocity rescaling. After reaching  equilibrium, energy, pressure, chain
radii of gyration, and end-to-end distances show no drift. Likewise, dynamic correlators show no aging effects.
Once the system is equilibrated, a microcanonical run is performed for production of configurations, from which static
and dynamic correlators are computed. Static correlators presented here are averaged over typically 
300 equispaced configurations. Dynamic correlators are averaged over typically 40 equispaced time origins. 
The typical duration of a production run is of 40-200 million time steps for respectively high and low temperatures.

\begin{center}
{\bf III. STATIC PROPERTIES}
\end{center}

\begin{center}
{\bf a)Orientational correlations}
\end{center}

\begin{figure}
\begin{center}
\includegraphics[width=0.85\linewidth]{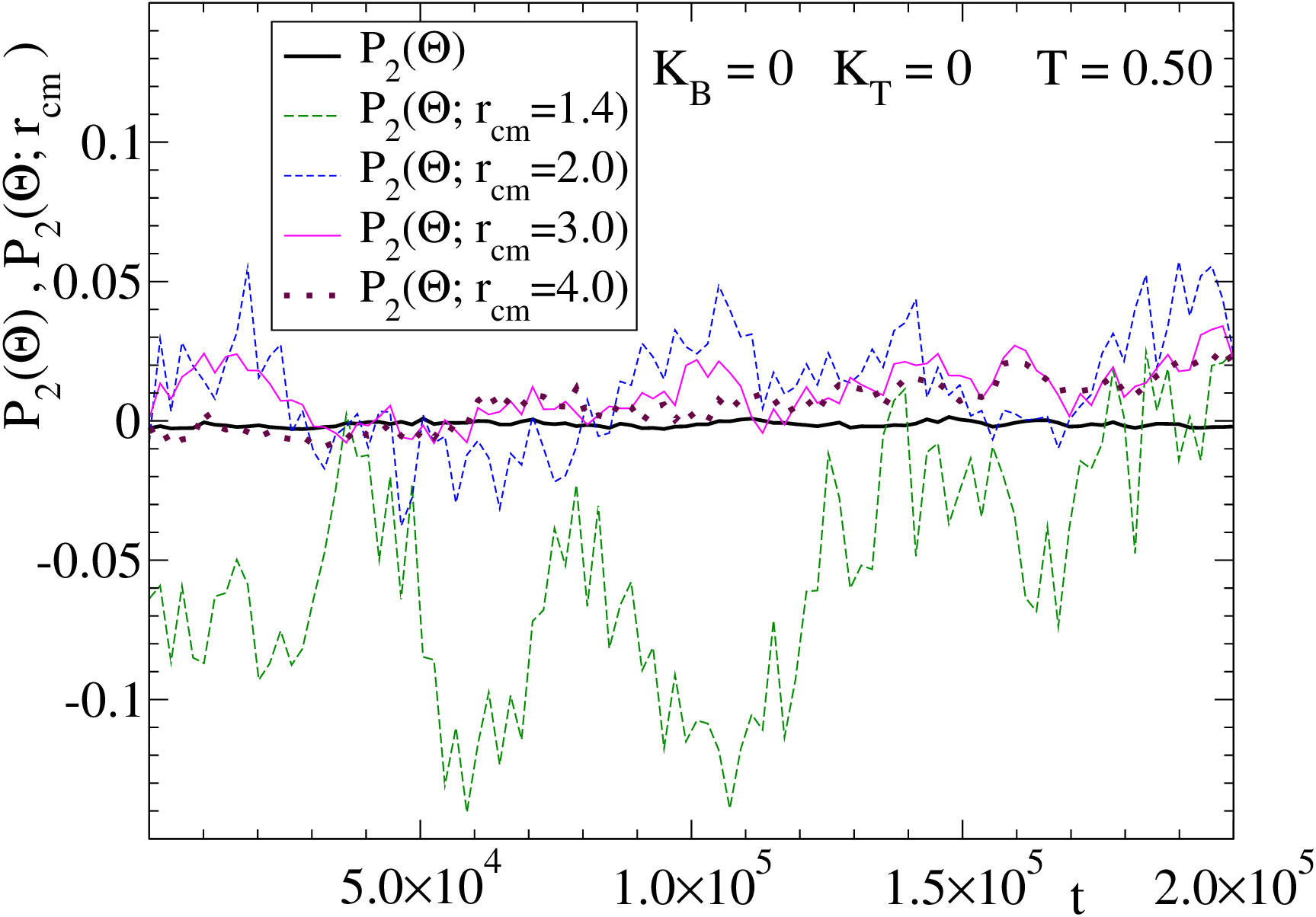}
\newline
\newline
\includegraphics[width=0.85\linewidth]{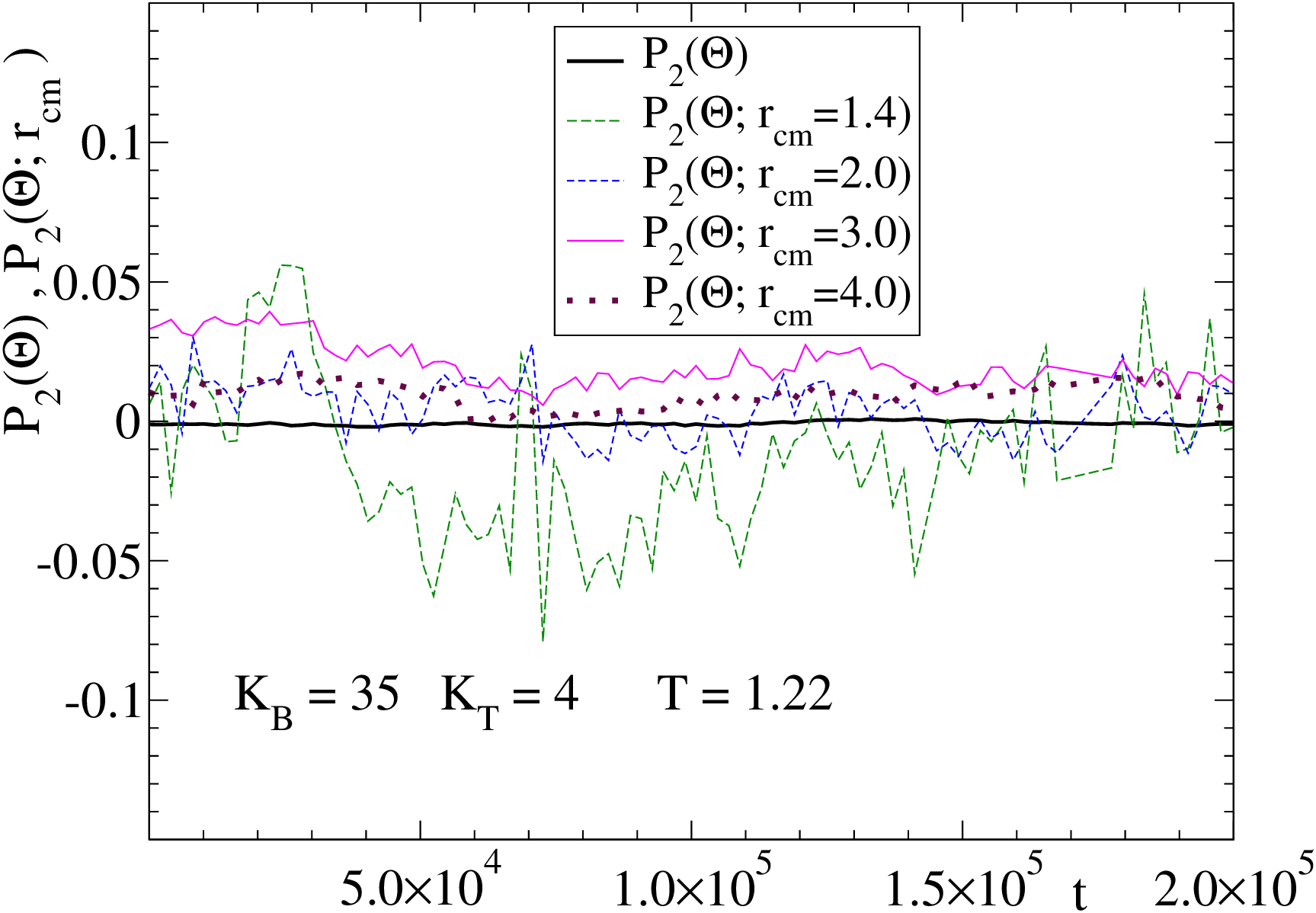}
\newline
\caption{Time evolution of the global and local
orientational parameter (see text) for fully-flexible (top) and stiff chains with 
$(K_{\rm B}$,$K_{\rm T}) =(35,4)$ (bottom), for two selected low temperatures.}
\label{fig:nematic}
\end{center}
\end{figure}

Simulation results presented in this work correspond to isotropic phases.
We do not observe signatures of {\it global} orientational order induced by chain stiffness
for the investigated state points. 
Thus, by measuring the quantity  $P_2 (\Theta) = (3\langle\cos ^2\Theta\rangle -1)/2$, 
where $\Theta$ is the angle between the end-to-end vectors of two chains, and averaging it 
over all pairs of distinct chains,  we obtain in all cases values $|P_2 (\Theta)| < 3\times 10^{-3}$.
This is illustrated in Fig. \ref{fig:nematic}, which shows the time evolution of $P_2 (\Theta)$ 
along a typical simulation window,
both for fully-flexible chains, $(K_{\rm B}$,$K_{\rm T}) =$ (0,0), 
and for representative stiff chains, $(K_{\rm B}$,$K_{\rm T}) =$ (35,4).

{\it Local} orientational order is also negligible. This is evidenced by computing
a similar correlator $P_2 (\Theta ;r_{\rm cm})$. In this case the average is performed only 
over pairs of distinct chains for which the distance between their respective 
centers-of-mass is less than $r_{\rm cm}$.
Fig. \ref{fig:nematic} displays, for the former cases of fully-flexible and stiff chains, 
data of $P_2 (\Theta ;r_{\rm cm})$ for several values of $r_{\rm cm}$.
Negligible values of $P_2 (\Theta ;r_{\rm cm})$ are obtained for $r_{\rm cm} \geq 2.0$. 
Thus, the time average
over the simulation time window, $t_{\rm sim}$, provides values 
$|\langle P_2 (\Theta ;r_{\rm cm} \geq 2.0)\rangle_{\rm t_{\rm sim}}| < 0.02$.
By comparing both panels we conclude that chain stiffness does not induce a significant 
increase, if any, of local orientational order. Weak local orientational order 
$|\langle P_2 (\Theta ;r_{\rm cm})\rangle_{\rm t_{\rm sim}}| \lesssim 0.1$ is observed
only for very small interchain distances (see data for $P_2 (\Theta ;r_{\rm cm} = 1.4)$). Again, the introduction
of chain stiffness does not induce clear changes in the orientational order at this length scale.
Note that for small $r_{\rm cm}$ differences in the represented data 
for different barrier strength may even be statistical artifacts, 
arising from the small number of neighboring chains within such distances and the limitted time of observation
(see the amplitude of the fluctuations in data of Fig. \ref{fig:nematic}).

\begin{center}
{\bf b)Static structure factors and chain form factors}
\end{center}

Now we present results for static structure factors and chain form factors,
both for fully-flexible chains and for a representative case of stiff chains. 
Let us consider an isotropic homogeneous system of volume $V$ containing $n$ identical chains of $N$ monomers.
The densities of chains and monomers are respectively denoted by $\rho_c=n/V$ and $\rho=nN/V$. 
Let us denote the location of a monomer along its chain by the index $1 \leq a \leq N$.
The site-site static structure factor for monomers of indices $a$ and $b$ is defined as:
\begin{equation}
S_{ab}(q)=\frac{1}{n}\langle\rho_a({\bf -q},0)\rho_b({\bf q},0)\rangle .
\label{eq:sqab}
\end{equation}
Brackets denote ensemble average.
The monomer density distribution for wave vector  ${\bf q}$ is given by
\begin{equation}
\rho_a({\bf q})=\sum_{j=1}^{n} \exp[i{\bf q}\cdot{\bf r}_{j}^{a}].
\label{eq:rhoa}
\end{equation}
In this expresion ${\bf r}_{j}^{a}$ is the position vector of the $a$th monomer
in the $j$th chain ($1 \leq j \leq n$). 
The quantity  $S_{ab}(q)$ can be splitted into intrachain 
and interchain $a$-$b$ correlations:
\begin{equation}
S_{ab}(q)=\omega_{ab}(q)+\rho_ch_{ab}(q),
\label{eq:sqabcompon}
\end{equation}
or in matrix form,
%
${\bf S}(q)= {\bf w}(q)+\rho_c {\bf h}(q)$.
%
In Eq. (\ref{eq:sqabcompon})  $\omega_{ab}(q)$ and $h_{ab}(q)$ respectively 
denote the intrachain and interchain correlations between monomers of type $a$ and $b$.
By averaging over all the possible pairs $(a,b)$ we obtain the static correlators $S(q)$, $\omega (q)$
and $h(q)$, which are related through:
\begin{equation}
S(q)= \omega(q) +\rho h(q).
\label{eq:Sq}
\end{equation}
In this expression $S(q)$ is the total static structure factor,
which equivalently can be obtained as  $S(q) = (nN)^{-1}\langle\rho(-q,0)\rho(q,0)\rangle$,
where $\rho(q)=\sum_{a=1}^{N} \rho_a(q)$ is the total monomer density distribution.
In Eq. (\ref{eq:Sq}) the chain form factor, $\omega (q)$, accounts
for all the static intrachain correlations, while $h(q)$ accounts for all the static interchain correlations.

\begin{figure}
\begin{center}
\includegraphics[width=0.84\linewidth]{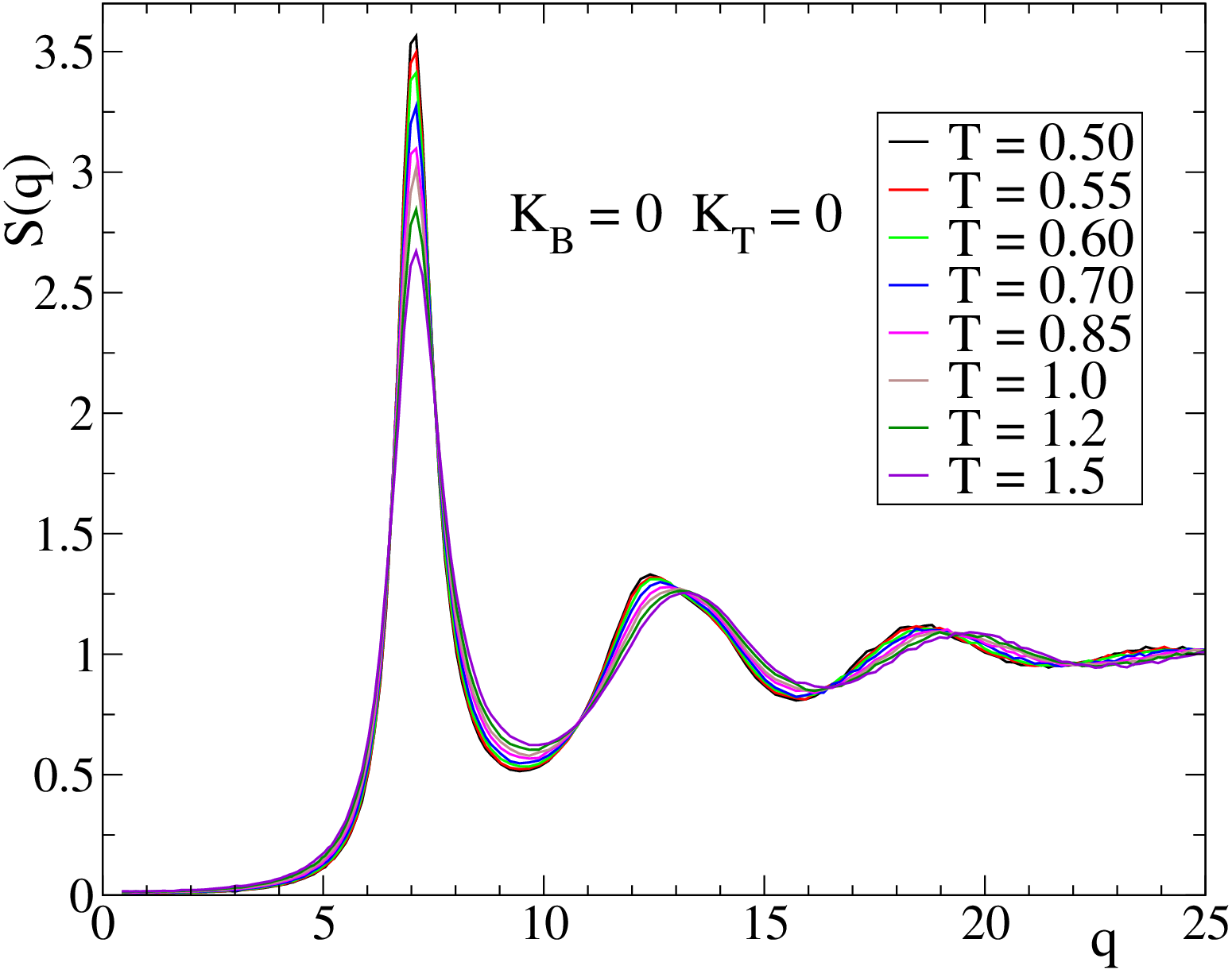}
\newline
\newline
\includegraphics[width=0.84\linewidth]{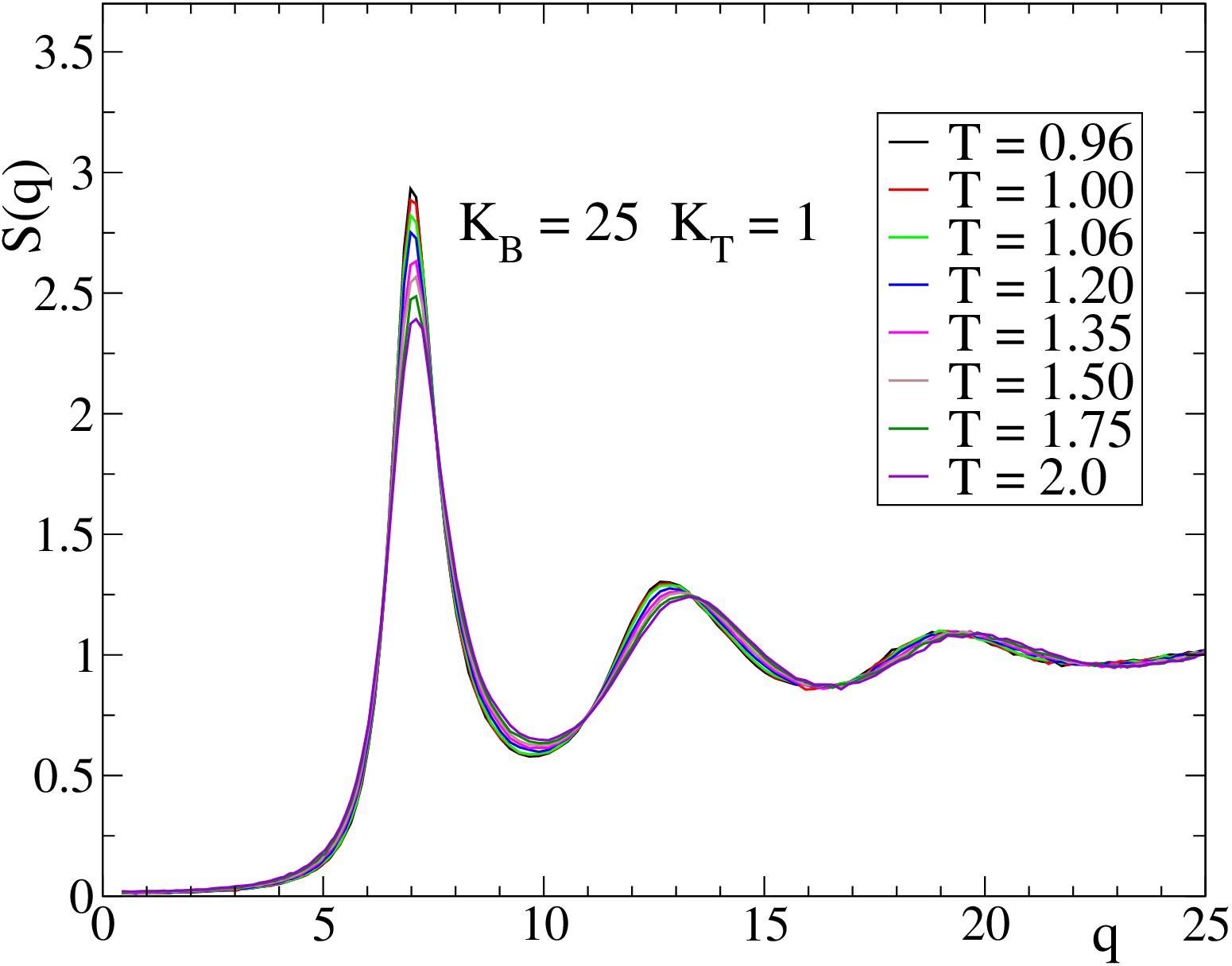}
\newline
\caption{Temperature dependence of the static structure factor $S(q)$ for fully-flexible chains
(top panel) and for chains with barrier strength $(K_{\rm B}, K_{\rm T}) = (25,1)$ (bottom panel).}
\label{fig:sqcomp}
\end{center}
\end{figure}

Fig. \ref{fig:sqcomp} (top panel) shows simulation results for $S(q)$ as a function of temperature
for fully-flexible chains, $(K_{\rm B}, K_{\rm T}) = (0,0)$. Data for representative stiff chains, 
$(K_{\rm B}, K_{\rm T}) = (25,1)$, are shown in the bottom panel. 
In both cases, no signature of crystallization is present. Indeed no sharp Bragg peaks are observed. 
In both cases $S(q)$ shows a maximum at $q_{\rm max}\approx 7.0$. Since $S(q_{\rm max})$ comes 
from the packing in the first shell around a monomer, the latter corresponds to a typical distance 
$2\pi/7.0 \approx 0.90$ in the real space between neighboring
monomers. On cooling, the peak at  $q_{\rm max}\approx 7.0$ increases in intensity, which is a signature of 
increasing short-range order.

\begin{figure}
\begin{center}
\includegraphics[width=0.89\linewidth]{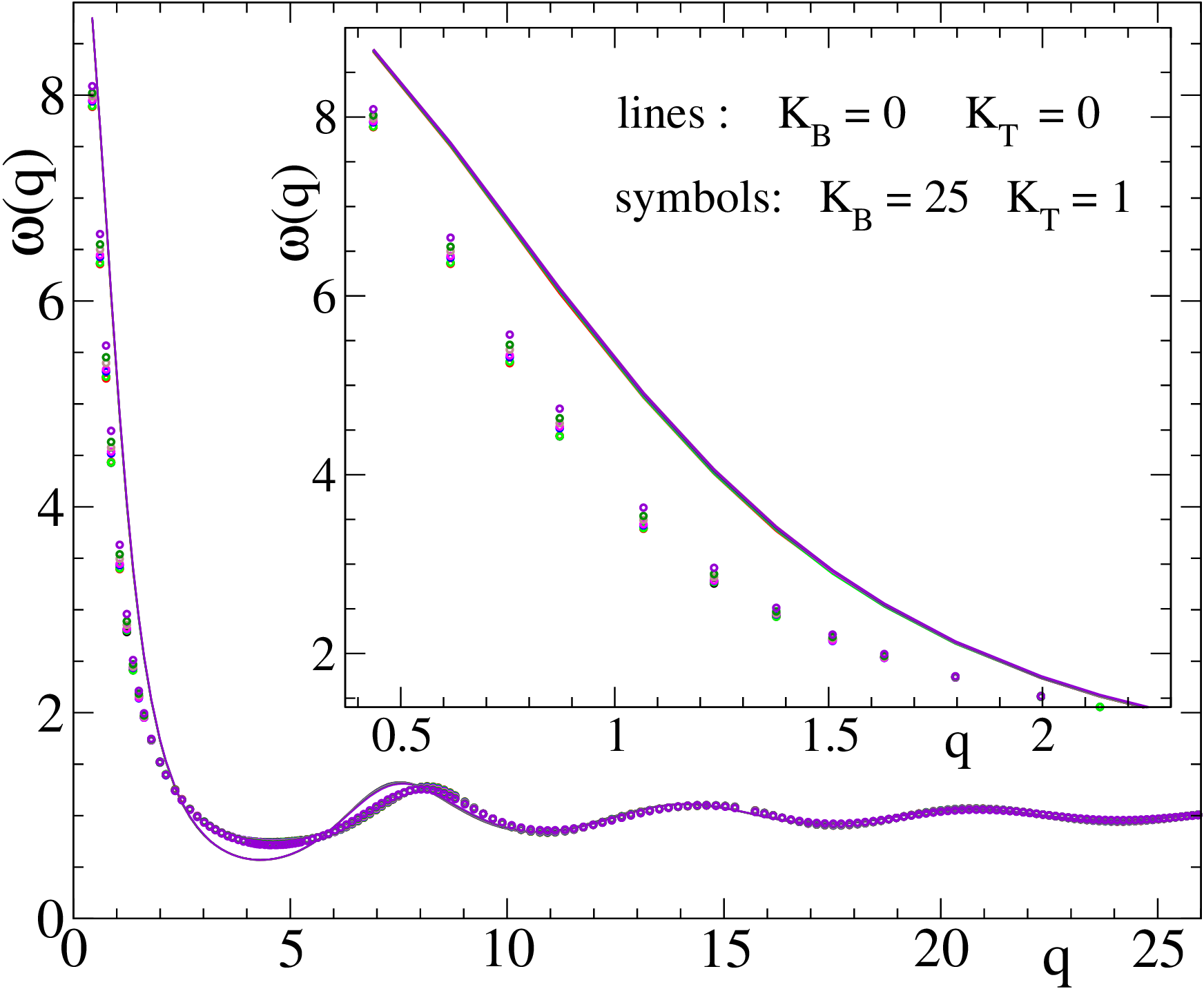}
\caption{Temperature dependence of the form factor $\omega(q)$ for fully-flexible chains (lines) and 
for chains with barrier strength $(K_{\rm B}, K_{\rm T}) = (25,1)$ (symbols).
For clarity, the inset shows results in the range of low-$q$. 
Different colors correspond to different temperatures, following the 
legends of Fig. \ref{fig:sqcomp}.}
\label{fig:wqcomp}
\end{center}
\end{figure}

In Fig.  \ref{fig:wqcomp} we show, for the former values of $(K_{\rm B}, K_{\rm T})$,
the corresponding results for the form factors $\omega(q)$. We note that in the case of fully-flexible
chains the form factor is nearly independent on temperature. 
The form factor for stiff chains
exhibits a certain $T$-dependence, which is however rather weak 
in comparison with that of $S(q)$.
The $T$-dependence of $\omega(q)$ becomes more clear at low $q$-values. 
The way the form factor behaves on lowering the temperature is directly connected with the values 
of the mean chain end-to-end radius $R_{\rm ee}$.
Thus, by decreasing temperature from $T = 2.0$ to $T= 0.96$, the computed $R_{\rm ee}$ increases 
from 4.8 to 5.5 for the selected stiff chains. This leads, for lower $T$, to a stronger decay 
in $\omega(q)$ at low-$q$. On the other hand, the value $R_{\rm ee}=3.6$ for the fully-flexible chains
is almost $T$-independent, leading to a negligible $T$-dependence of $\omega(q)$.

\begin{center}
{\bf c)Test of the PRISM approximations}
\end{center}
The MCT for polymer melts developed by Chong and co-workers \cite{chongprl,chongpre} invokes 
several approximations of the PRISM theory \cite{curro}. In this subsection we summarize
such approximations and test their validity for all the investigated range of barrier strength.
The site-site direct correlation function, $c_{ab}(q)$,
is introduced via the generalized Ornstein-Zernike relation for polyatomic molecules,
or `reference interaction site model' (RISM) \cite{rism},
\begin{equation}
h_{ab}(q)=\sum_{x,y=1}^{N} \omega_{ax}(q)c_{xy}(q) [\omega_{yb}(q)+\rho_c h_{yb}(q)],
\label{eq:rism}
\end{equation}
in which intramolecular contributions are accounted by the form factor terms $\omega_{ab}(q)$.
By inserting (\ref{eq:sqabcompon}) in Eq. (\ref{eq:rism}),  $c_{ab}(q)$ is related 
to $S_{\rm ab}(q)$ and $\omega_{ab}(q)$ as:
\begin{equation}
\rho_{\rm c} c_{ab}(q)= \omega_{ab}^{-1}(q)-S_{ab}^{-1}(q) .
\label{eq:cabq}
\end{equation}
Here $\omega_{ab}^{-1}(q)$ and $S_{ab}^{-1}(q)$ are the elements of, respectively,
the matrices ${\bf w}^{-1}(q)$ and ${\bf S}^{-1}(q)$,
which are defined as the inverses of ${\bf w}(q)$ and ${\bf S}(q)$.

In the {\it equivalent-site approximation} (which is exact for polymer rings) of the PRISM,
chain end effects are neglected and all sites are treated equivalently
for interchain correlations. Thus, $c_{ab}$ is replaced by the average over all $(a,b)$-pairs: 
\begin{equation}
c_{ab}(q) \approx c(q) .
\label{eq:equivsite}
\end{equation}
By introducing this approximation in Eq. (\ref{eq:rism}) and averaging 
over all $(a,b)$-pairs we find 
$h(q) = \omega(q)c(q) [\omega(q)+\rho h(q)]$. By introducing Eq. (\ref{eq:Sq})
in the latter expression we arrive to the scalar equation
\begin{equation}
\rho c(q) = 1/\omega(q)-1/S(q) ,
\label{eq:prism}
\end{equation}
also known as PRISM equation.

\begin{figure}
\begin{center}
\includegraphics[width=0.8\linewidth]{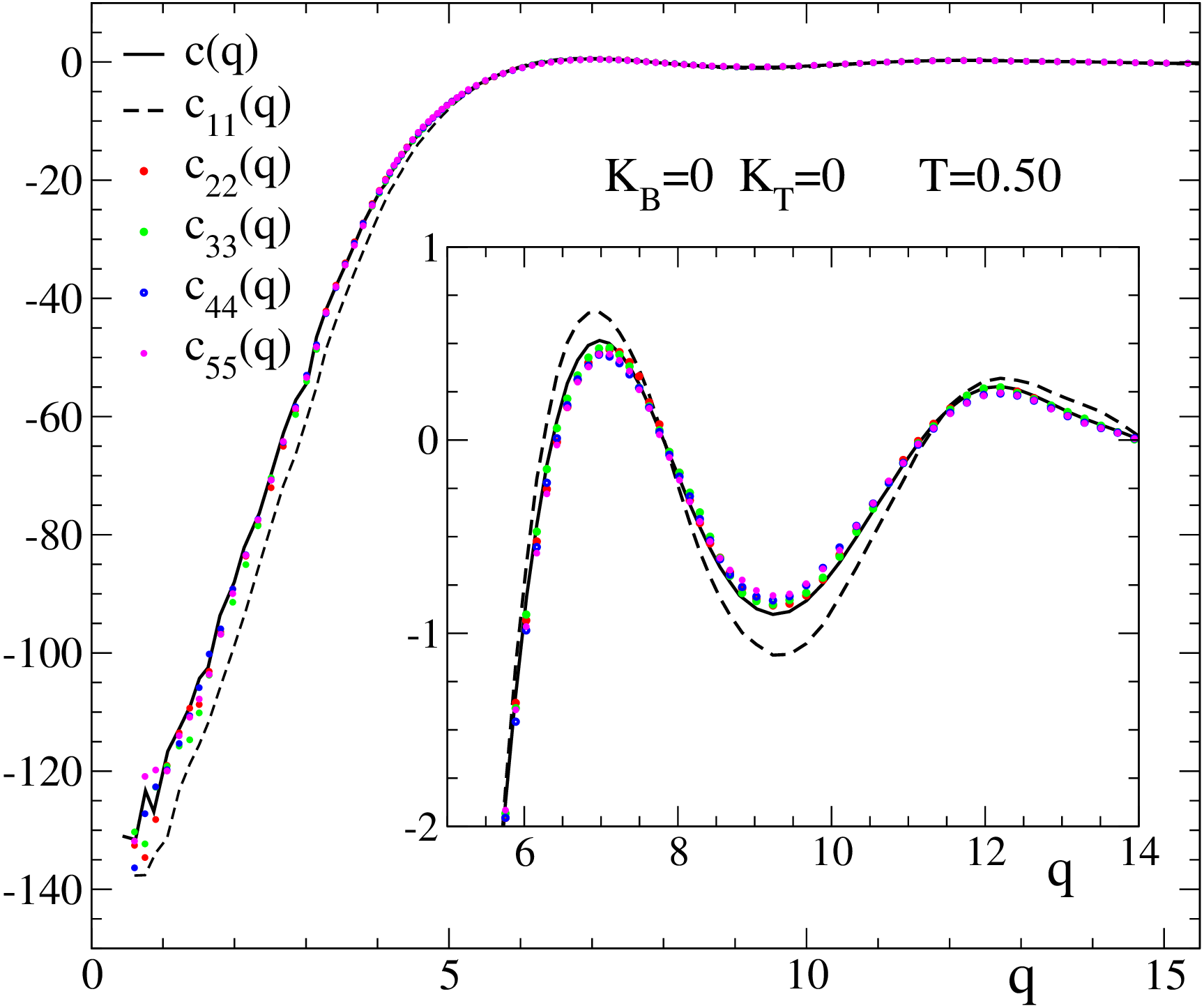} 
\newline
\newline
\includegraphics[width=0.8\linewidth]{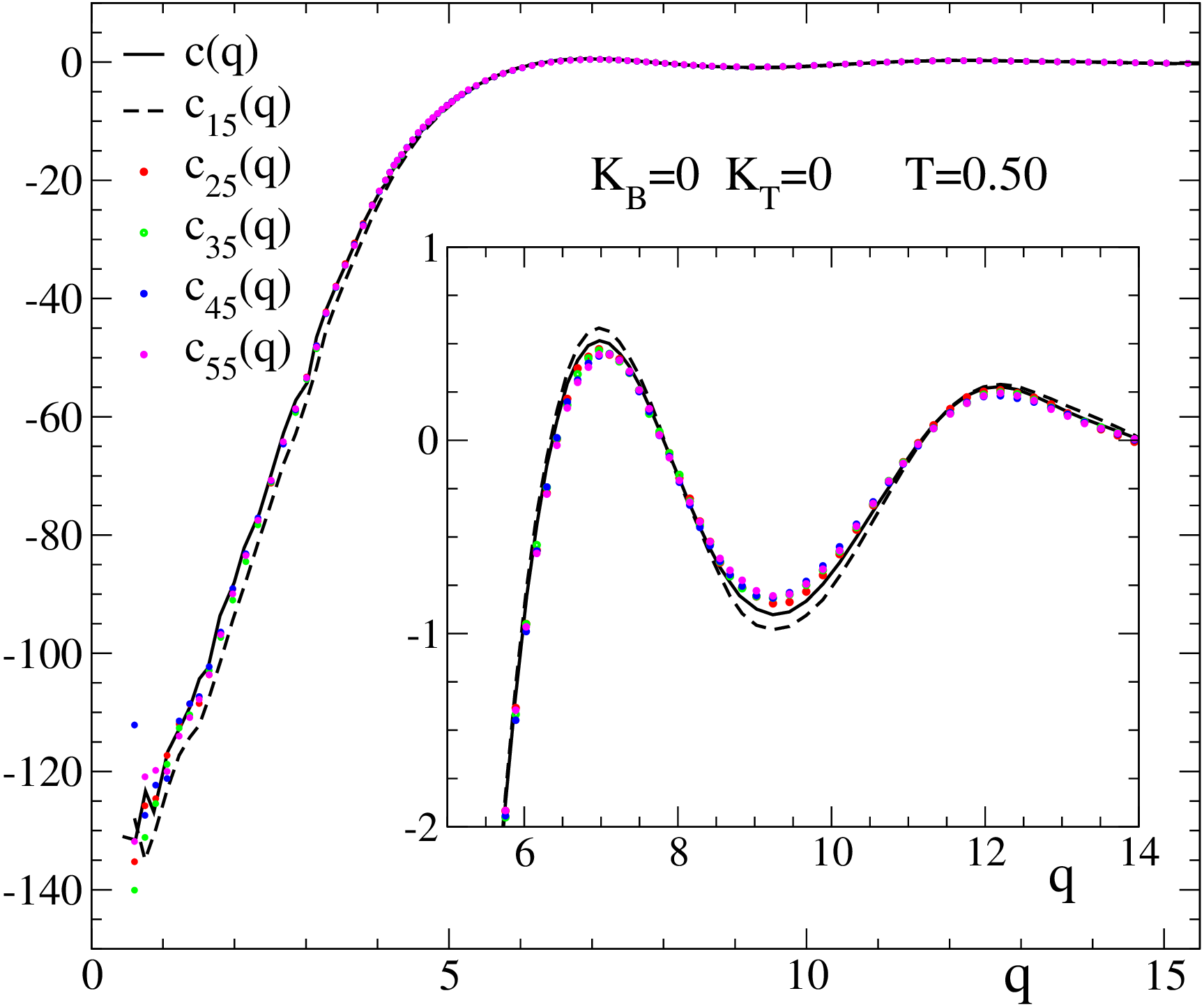}
\newline
\caption{Test of the equivalent site approximation, Eq. (\ref{eq:equivsite}), 
for fully-flexible chains at  $T=0.50$.
Top and bottom panels compare $c(q)$ with respectively matrix elements $c_{aa}(q)$ and $c_{a5}(q)$.
The insets enhance the region around the wave vector $q_{\rm max}$ 
for the maximum of the static structure factor $S(q)$.}
\label{fig:eqsite}
\end{center}
\end{figure}

\begin{figure}
\begin{center}
\includegraphics[width=0.8\linewidth]{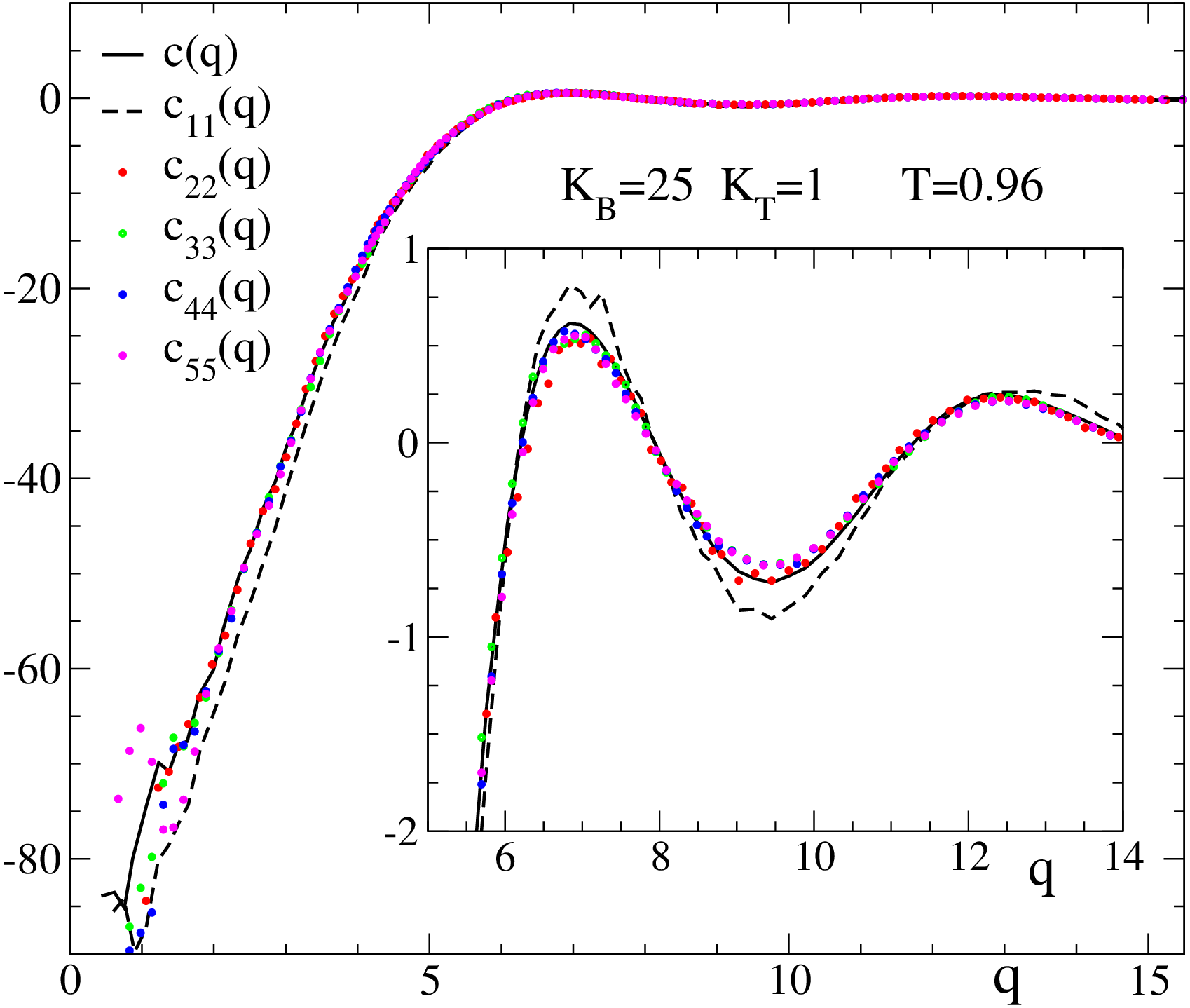} 
\newline
\newline
\includegraphics[width=0.8\linewidth]{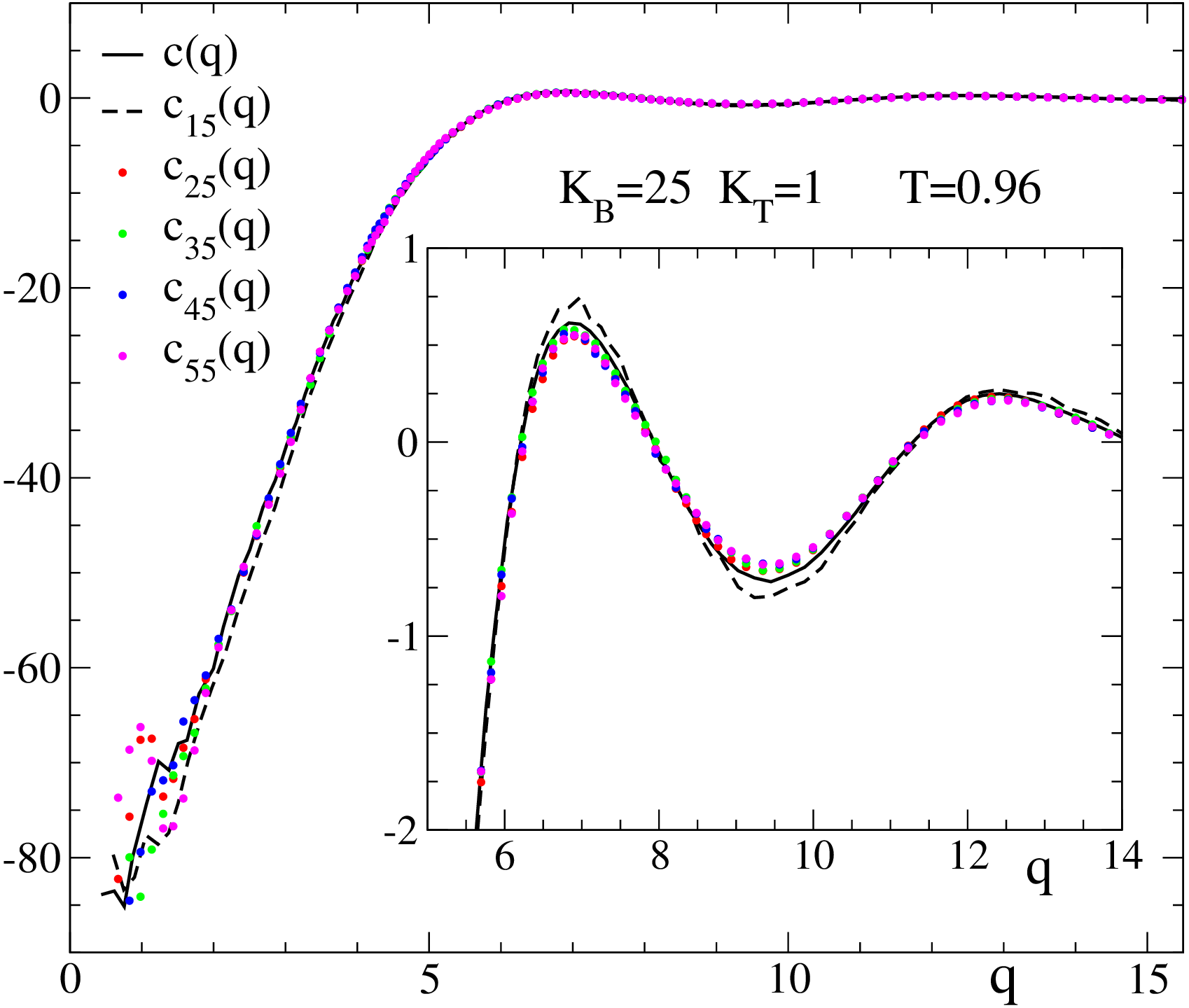}
\newline
\caption{Test of the equivalent site approximation, Eq. (\ref{eq:equivsite}), 
for stiff chains with $(K_{\rm B}, K_{\rm T}) = (25,1)$, at  $T=0.96$.
Top and bottom panels compare $c(q)$ with respectively matrix elements $c_{aa}(q)$ and $c_{a5}(q)$.
The insets enhance the region around the wave vector $q_{\rm max}$ 
for the maximum of the static structure factor $S(q)$.}
\label{fig:eqsite2}
\end{center}
\end{figure}

In Figs. \ref{fig:eqsite} and \ref{fig:eqsite2} we test the validity 
of the equivalent-site approximation $c_{ab}(q) \approx c(q)$. We calculate 
$c_{ab}(q)$ and $c(q)$ respectively through Eqs. (\ref{eq:cabq}) and (\ref{eq:prism}),
by using the quantities $\omega_{ab}^{-1}(q)$, $S_{ab}^{-1}(q)$, $\omega (q)$, and $S(q)$ 
as computed from the simulations. Fig. \ref{fig:eqsite} shows results for
the fully-flexible case. Data for the case $(K_{\rm B}, K_{\rm T}) = (25,1)$
are displayed in Fig. \ref{fig:eqsite2}. Both data sets correspond to the respective
lowest investigated temperatures. We use a representation analogous
to that of Ref. \cite{aichelepre}. Thus, top and bottom panels in both figures
show the comparison of the averaged $c(q)$  with respectively the matrix elements
$c_{aa}(q)$ and $c_{a5}(q)$. The data of Fig.  \ref{fig:eqsite}  are consistent
with results of Ref. \cite{aichelepre} for a similar fully-flexible bead-spring model.
Data in Fig. \ref{fig:eqsite2} constitute new results for the case of implemented 
intramolecular barriers. By looking at both figures we conclude that the quality
of the equivalent-site approximation is not altered by the introduction
of strong intramolecular barriers. Data in Fig. \ref{fig:eqsite2} 
display the same trends as in the fully-flexible case. Thus, $c_{ab}(q) \approx c(q)$
is an excellent approximation except for correlations involving chain end monomers $a=1$
(and $a=N$ by symmetry). The latter show deviations from $c(q)$ which are moderate
around $q_{\rm max}$, this $q$-range being the dominating one in the MCT kernel.

\begin{figure}
\begin{center}
\includegraphics[width=0.8\linewidth]{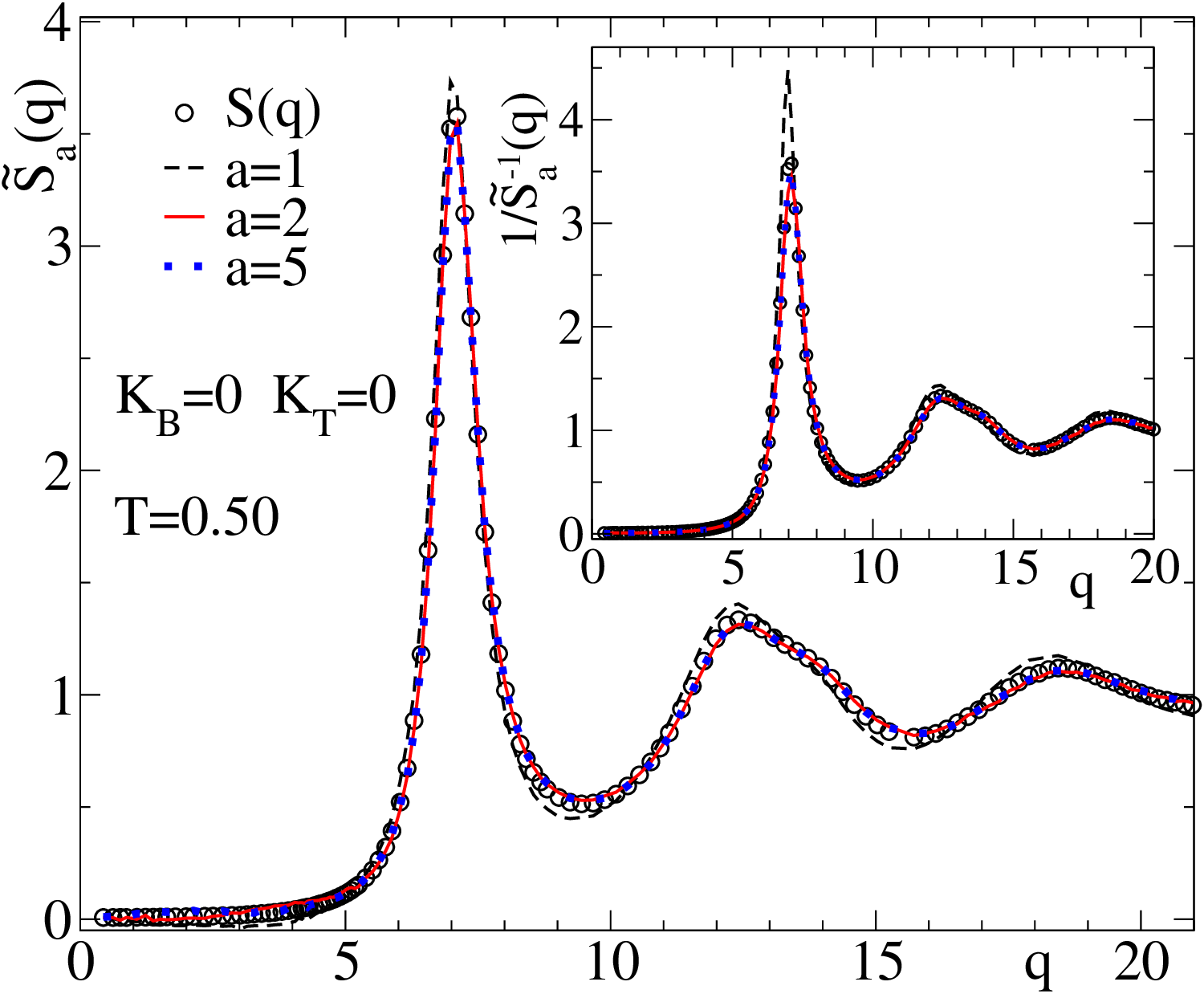} 
\newline
\newline
\includegraphics[width=0.8\linewidth]{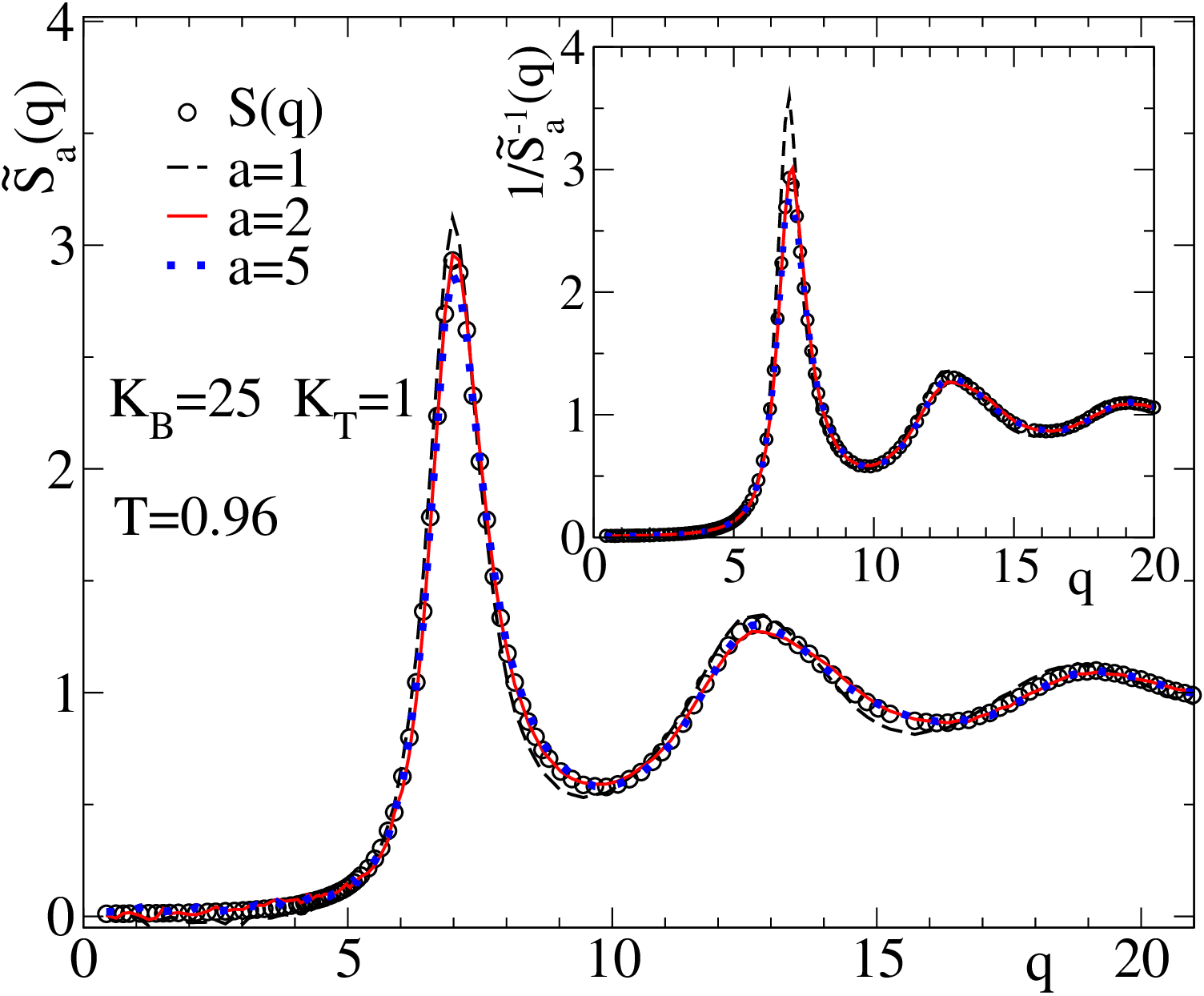} 
\newline
\caption{Test of the ring approximation, Eqs. (\ref{eq:ring},\ref{eq:invring}). 
Top panel: fully-flexible case at $T=0.50$.
Bottom panel: barrier strength $(K_{\rm B}, K_{\rm T}) = (25,1)$, at  $T=0.96$.  
Main panels and insets compare $S(q)$ (symbols) with respectively  
$ \tilde{S}_a(q)$ and 1/$\tilde{S}_a^{-1}(q)$, for the sites $a=$ 1,2 and 5 (lines).}
\label{fig:ring}
\end{center}
\end{figure}

An additional approximation of the PRISM is the {\it ring approximation}
(which is again exact for polymer rings). First we define the quantities
$\tilde {S}_a(q) = \sum_{b=1}^{N}S_{ab}(q)$ and $\tilde{S}_a^{-1}(q) = \sum_{b=1}^{N}S_{ab}^{-1}(q)$. 
By exploiting the fact that
for a ring polymer  $\tilde{S}_a(q)$ is $a$-independent, i.e., 
$\tilde{S}_a(q) \approx N^{-1}\sum_{a=1}^N \tilde{S}_a (q)$, we find 
\begin{equation}
\tilde{S}_a (q)  \approx S(q) .
\label{eq:ring}
\end{equation}
From the definition of $S_{ab}^{-1}(q)$ and $\tilde{S}_b(q)$ 
the relation $\sum_{b=1}^N S_{ab}^{-1}(q) \tilde{S}_b(q) =1$  is exact.
By introducing the ring approximation $\tilde{S}_b(q) \approx S(q)$ 
the former relation is transformed into:
\begin{equation}
\frac{1}{\tilde{S}_a^{-1}(q)}   \approx S(q)  .
\label{eq:invring}
\end{equation}
Fig. \ref{fig:ring} shows a test of the ring approximation of 
Eqs. (\ref{eq:ring}) (main panels) and (\ref{eq:invring}) (insets). This is done both for fully-flexible
chains (top panel) and for stiff chains with $(K_{\rm B}, K_{\rm T}) = (25,1)$ (bottom panel).
The comparison between $S(q)$, $\tilde{S}_a (q)$ and $1/\tilde{S}_a^{-1} (q)$ as computed from simulations
is in general excellent, with the same quality for fully-flexible and stiff chains. 
Only for the end monomers $a=1$ (and $a=N$ by symmetry) significant differences 
between $S(q)$ and $1/\tilde{S}_a^{-1} (q)$
are observed around the wave vector $q_{\rm max}$.

With all these results we conclude that the approximations assumed by the PRISM theory and introduced 
in the MCT equations for polymer melts (see below) are fulfilled, with the same quality for all the investigated
range of barrier strength.

\begin{center}
{\bf IV. DYNAMIC PROPERTIES}
\end{center}

\begin{figure}
\begin{center}
\includegraphics[width=0.9\linewidth]{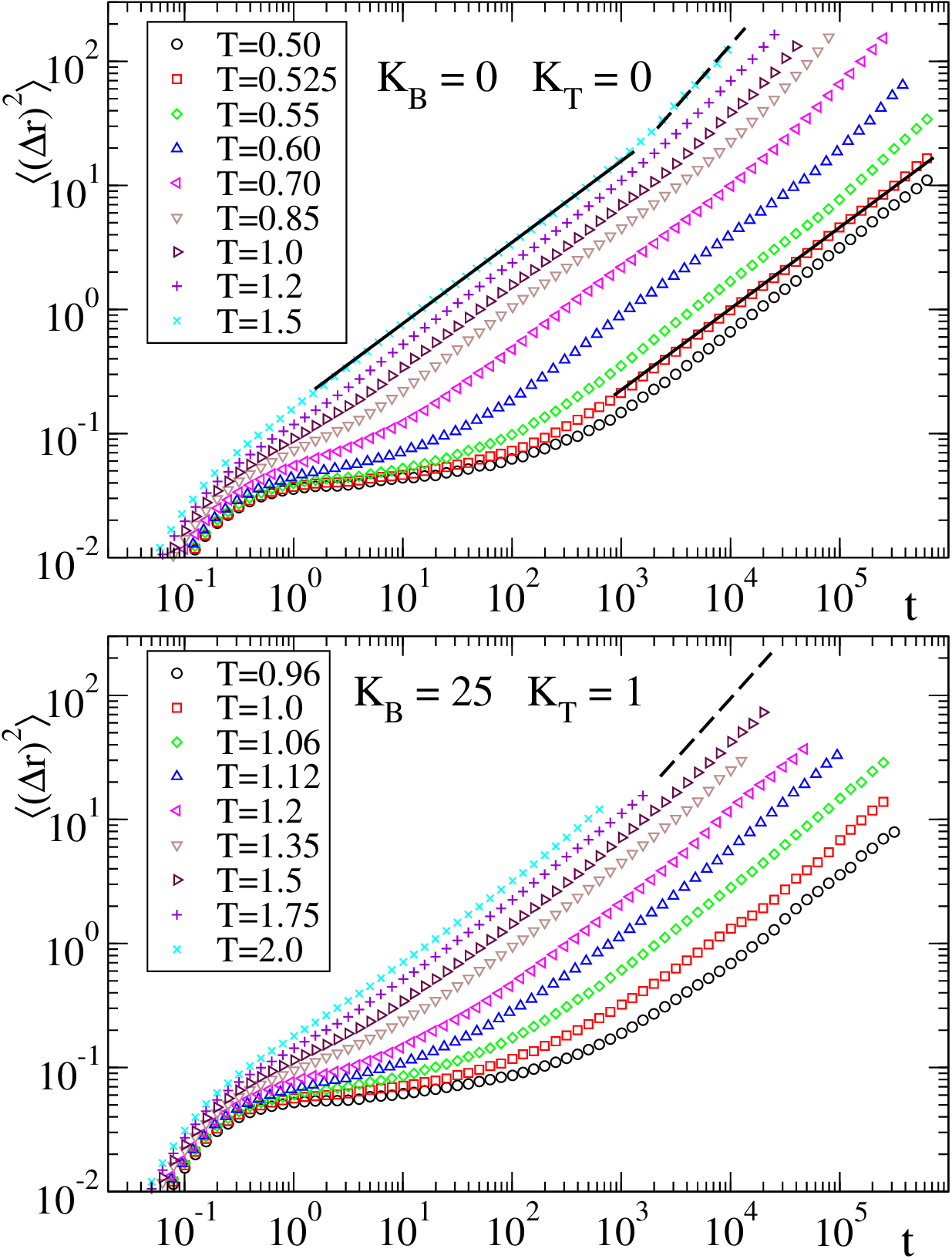}
\caption{Temperature dependence of the monomer mean squared displacement 
for fully-flexible (top) and stiff chains with $(K_{\rm B}, K_{\rm T}) = (25,1)$
(bottom). The solid and dashed lines indicate respectively sublinear ($\sim t^{0.65}$)
and linear behavior.}
\label{fig:msdKT0KT25}
\end{center}
\end{figure}

In this section we show some phenomenological dynamic features
induced by the introduction of intramolecular barriers in our model.
Panels in Fig. \ref{fig:msdKT0KT25} show the $T$-dependence of the monomer 
mean squared displacement (MSD) for fully-flexible 
and representative stiff chains with $(K_{\rm B}, K_{\rm T}) = (25,1)$. 
We observe similar features in both cases, but also some differences. 
After the initial ballistic regime, 
a plateau extends over longer times with decreasing temperature. This plateau corresponds 
to the caging  regime --- i.e., the temporary trapping of each
monomer in  the shell of neighboring  monomers around it --- which is usually observed
when approaching a liquid-glass transition.
At longer times, leaving the plateau, a crossover to a Rouse-like  
sublinear regime $\langle(\Delta r)^2\rangle \propto t^{0.65}$ \cite{bennemann,aichele}
is observed for the fully-flexible case. The final crossover to linear diffusion 
$\langle(\Delta r)^2\rangle \propto t$ is reached at long times 
only for the highest investigated temperatures.
However, for the case of stiff chains it is difficult to discriminate
power-law behavior over significant time windows. Apparently, the linear diffusive regime
is not reached within the simulation time window.

\begin{figure}
\begin{center}
\includegraphics[width=0.83\linewidth]{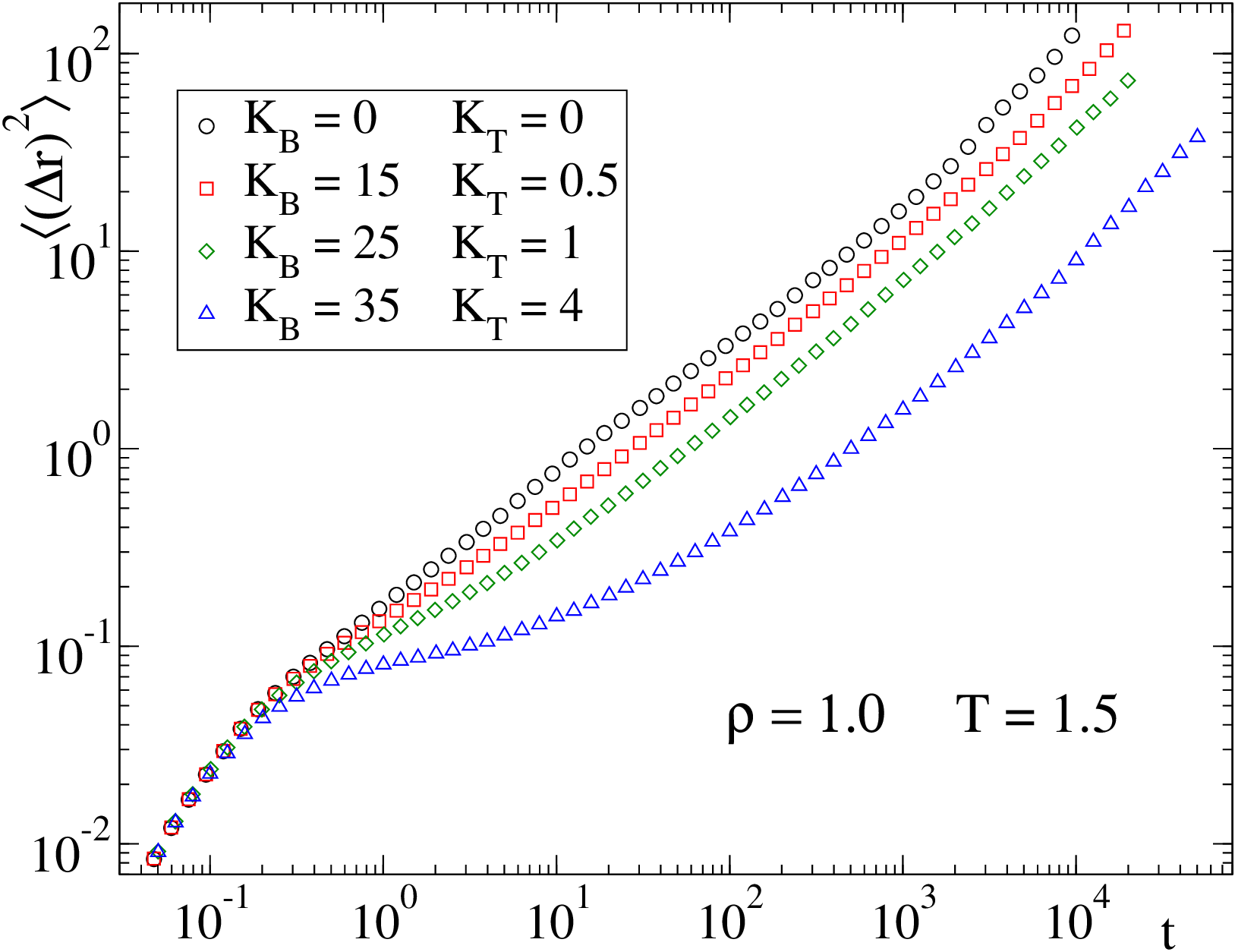}
\caption{Monomer mean squared displacement, for several values of the barrier strength,
at fixed density $\rho = 1.0$ and temperature $T = 1.5$.}
\label{fig:compmsd}
\end{center}
\end{figure}

Fig. \ref{fig:compmsd} shows the monomer  MSD, for fixed values of
density $\rho=1.0$ and temperature $T = 1.5$, as a function of the barrier strength. 
Consistently with results in Ref. \cite{bulacu2},
we observe that increasing the strenght of the internal barriers at fixed $\rho$ and $T$ 
leads to slower dynamics.

\begin{figure}
\begin{center}
\includegraphics[width=0.88\linewidth]{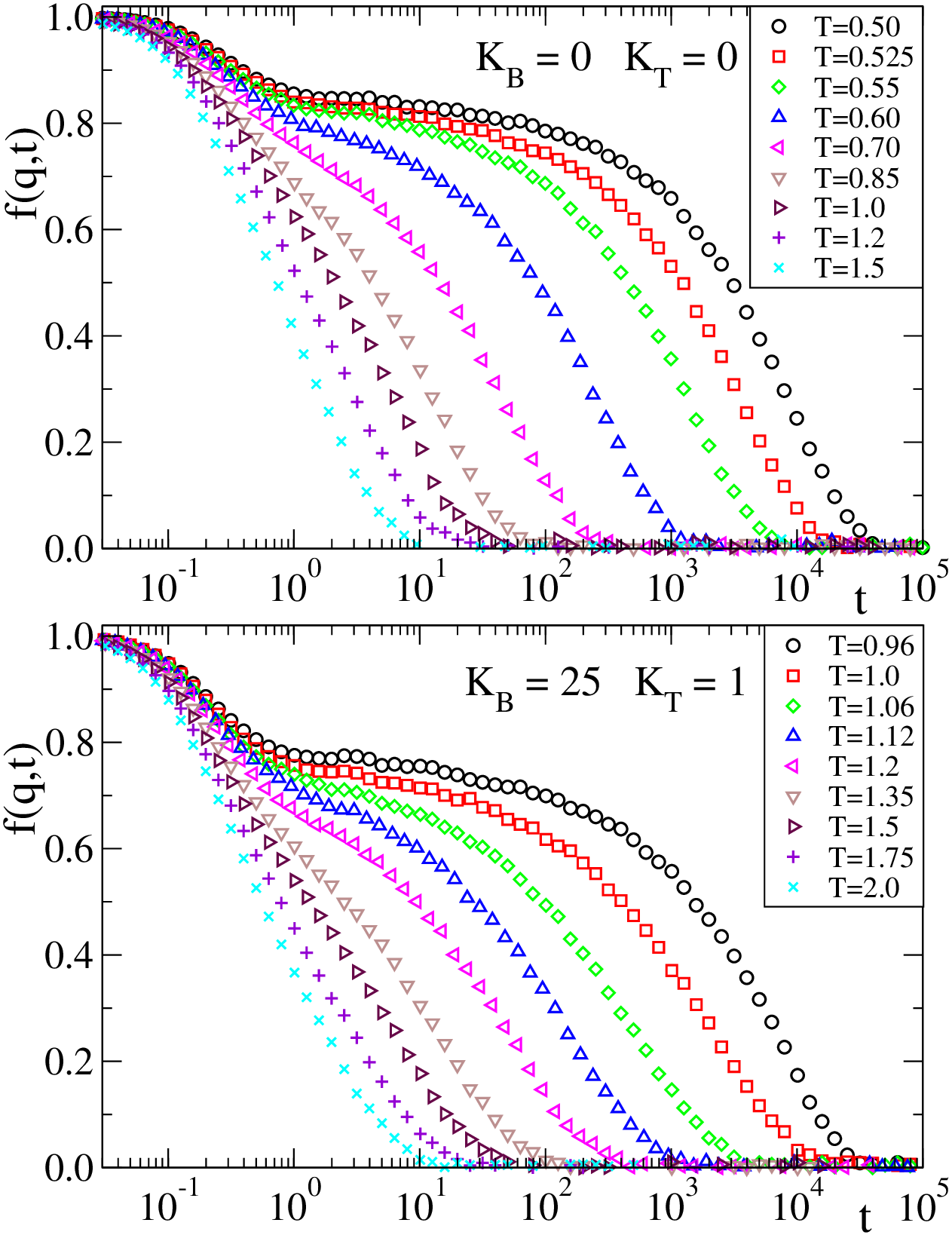}
\caption{Temperature dependence of $f(q,t)$ for fully-flexible (top) and stiff chains
with $(K_{\rm B}, K_{\rm T}) = (25,1)$ (bottom). 
The wave vector is $q_{\rm max} \approx7$, in both cases corresponding to the 
maximum of the static structure factor $S(q)$.}
\label{fig:compfcohqt}
\end{center}
\end{figure}

Fig. \ref{fig:compfcohqt} shows simulation results at several temperatures, 
both for fully-flexible and stiff chains, for the normalized density-density correlator $f(q,t)$. 
The latter is defined as $f(q,t) = \langle \rho(-q,0)\rho(q,t)\rangle/\langle \rho(-q,0)\rho(q,0)\rangle$.  
In both cases the correlator is evaluated at the 
maximum, $q_{\rm max} \approx 7$, of the static structure factor $S(q)$. 
As in the case of the MSD, both the  fully-flexible and stiff cases exhibit the standard behavior in the
proximity of a glass transition \cite{bennemann,aichele}. 
After the initial transient regime, $f(q,t)$ shows a first decay to a plateau connected 
with the caging regime. On lowering the temperature this plateau extends over longer time intervals. 
At long times, a second decay is observed from the plateau to zero. 
This second decay corresponds to the structural $\alpha$-relaxation. 

\begin{figure}
\begin{center}
\includegraphics[width=0.82\linewidth]{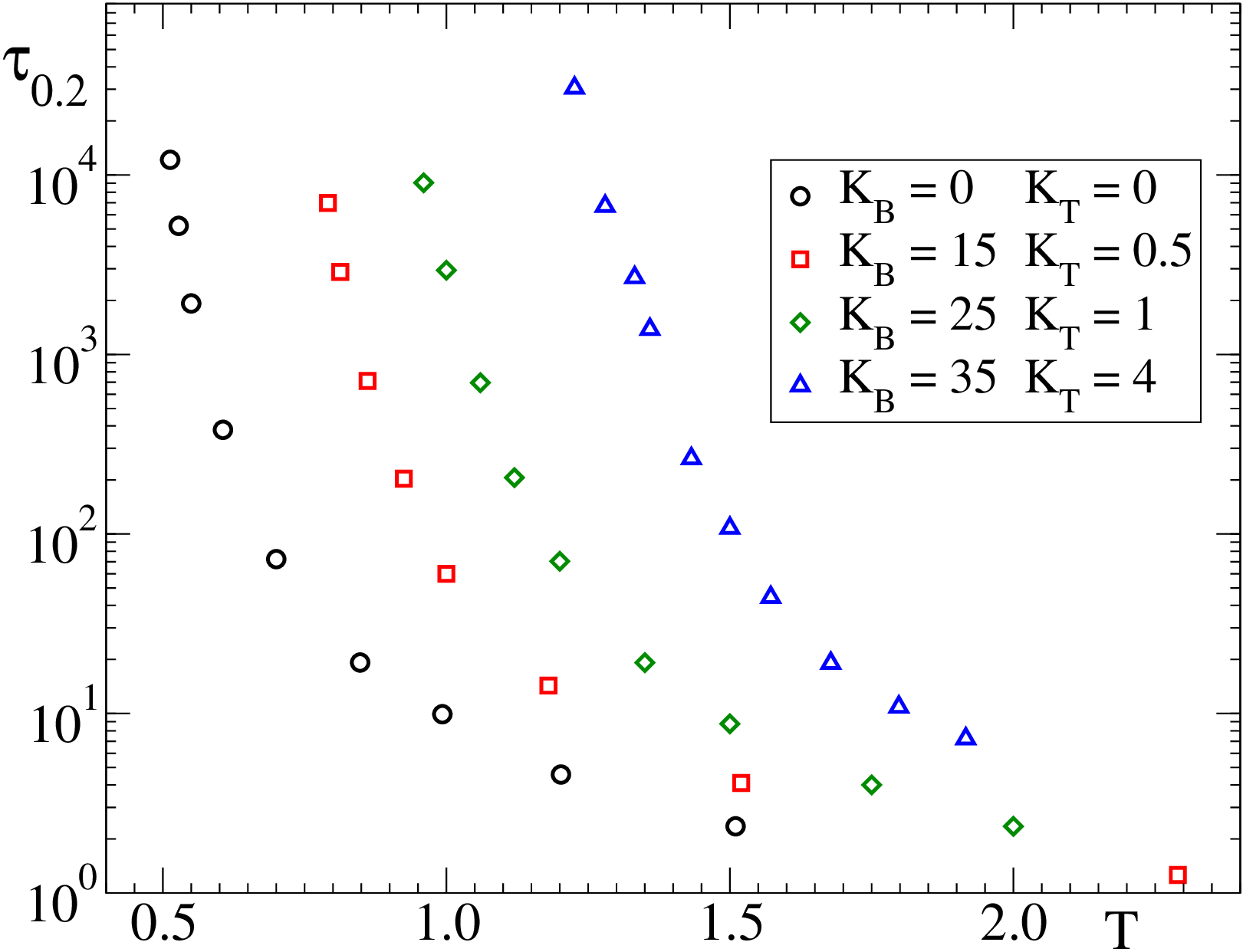}
\caption{Temperature dependence of the relaxation times $\tau_{0.2}$
for several values of the barrier strength.}
\label{fig:compreltime}
\end{center}
\end{figure}

Let us define the relaxation time as a time scale probing the $\alpha$-structural relaxation.
This can be done by introducing the time $\tau_x$ for which the correlator for $q_{\rm max}$
takes the value $f(q_{\rm max},\tau_x) = x$, provided $x$ is small in comparison with the plateau height. 
Here we use $x = 0.2$. Fig. \ref{fig:compreltime} shows $\tau_{0.2}$ as a funcion of $T$, 
for different values of the bending and torsional constants.
As observed in the analysis of the mean squared displacements, increasing the chain stiffness 
slows down the dynamics. At fixed temperature, the relaxation time for the stiffest 
investigated chains increases up to 
three decades with respect to the fully-flexible case. 

In this section we have demonstrated a main dynamic feature: 
the slowing down of the dynamics, at fixed density and temperature,
by progressively increasing the strength of the intramolecular barriers. 
This feature strongly suggests that intramolecular barriers constitute
and additional mechanism for dynamic arrest, coexisting with the
general packing effects induced by density and temperature.
In the following we summarize the main predictions of the Mode Coupling Theory
and discuss simulation dynamic features within this theoretical framework.

\begin{center}
{\bf V. MODE COUPLING THEORY: SUMMARY}
\end{center}
 
In this section we briefly summarize universal dynamic scaling laws concerning the 
MCT liquid-glass dynamics, and test them in the simulated polymer melt for all the
investigated range of barrier strength.
Extensive reviews on MCT can be found, e.g., 
in Refs. \cite{mctrev1,mctrev2,gotzebook,reichman,das,gotzenewbook,franosch,fuchspre}.
Though initially derived for simple hard-sphere systems, 
these predictions follow as consequences of the mathematical structure 
of the MCT equations. More specifically, they are associated to 
the bilinear dependence of the memory kernel on the density correlators (see below).
Thus, MCT predicts the same dynamic scaling laws of the monoatomic case 
if such a mathematical structure is retained in systems of polyatomic molecules.
This is indeed the case of the MCT for polymer melts developed by Chong
and co-workers \cite{chongprl,chongpre} (see below). Therefore, 
the phenomenological analysis of our simulation results
in terms of MCT dynamic scaling laws is justified within the theory.

By starting from the fundamental Liouville equation of motion and  
using the Mori-Zwanzig projection operator
formalism  one arrives to an integro-differential equation for the normalized 
density-density correlator: 
\begin{eqnarray}
\ddot{f}(q,t) + \frac{q^2k_BT}{mS(q)}  f(q,t) \hspace{2 cm} \nonumber \\
+\frac{q^2k_BT}{mS(q)}\int^{t}_{0}  dt' m(q,t-t') \dot{f}(q,t') = 0 .
\label{eq:fMCT}
\end{eqnarray}
This equation is obtained by using projectors over the subspace spanned by 
the densities and the longitudinal currents. The memory kernel
$m(q, t-t') \propto \langle R_q (0) R_q (t-t') \rangle$, where the quantities $R_q$ are,
within the Mori-Zwanzig formalism, the associated fluctuating forces.
Since the kernel cannot be exactly expressed
in terms of $f(q,t)$ and/or its time derivatives, Eq. (\ref{eq:fMCT}) is not solvable.
MCT introduces several approximations for the memory kernel, 
in order to provide a closed solvable form of Eq. (\ref{eq:fMCT}). These approximations are:
\\
\\
i) It is assumed that the long-time, slow dynamic regime of any observable coupled to density fluctuations
can be expressed as a linear combination of `mode pairs', $\rho_{{\bf k}}\rho_{{\bf q-k}}$.
Since the {\it exact} expression of the correlator of the fluctuating forces 
contains a slow contribution which is 
a linear combination of mode pairs (see e.g., Ref. \cite{reichman} for details), the former assumption
is equivalent to neglecting the fast contribution of the fluctuating forces. In other words,
it is equivalent to assuming a large separation between the time scales of the former contributions.
\\
\\
ii) Convolution approximation: three-point static correlations are approximated as products
of static structure factors, 
\begin{eqnarray}
\langle \rho_{-{\bf q}}(0) \rho_{{\bf k}}(0) \rho_{{\bf q}-{\bf k}}(0) \rangle 
\approx nN S(q)S(k)S(|{\bf q}-{\bf k}|) .
\label{eq:convol}
\end{eqnarray}
iii) Kawasaki approximation: dynamic four-point correlations are
factorized in terms of products of dynamic two-point correlations, 
\begin{eqnarray}
\langle \rho_{{\bf k}-{\bf q}}(0)\rho_{-{\bf q}}(0) \rho^{\cal{Q}}_{{\bf q}-{\bf k}}(t)\rho^{\cal{Q}}_{{\bf q}}(t) \rangle  \approx
\hspace{1 cm} \nonumber \\
F(q,t)F(|{\bf q}-{\bf k}|,t) ,
\label{3approxMCT}
\end{eqnarray}
where the superscript $\cal{Q}$ denotes evolution with projected dynamics (see e.g., Ref. \cite{reichman} for
details), and $F(q,t) = \langle \rho(-q,0) \rho(q,t) \rangle$ are just the unnormalized 
density-density correlators. 
\\
\\
By making use of these three approximations,  the memory kernel 
$m(q,t)$ becomes  a bilinear form in $f(q,t)$,
\begin{equation}
m(q,t) = \int\frac{d^3 {\bf k}}{(2\pi)^3}{\cal V}({\bf q}-{\bf k},
{\bf k})f(k,t)f(|{\bf q}-{\bf k}| ,t) ,
\label{eq:mMCT}
\end{equation}
where the vertex ${\cal V}({\bf q}-{\bf k},{\bf k})$ is given by:
\begin{eqnarray}
{\cal V}({\bf q}-{\bf k},{\bf k})= \frac{\rho}{2q^4} 
S(q)S(k)S(|{\bf q}-{\bf k}|) \times \nonumber \\
\left[{\bf q}\cdot{\bf k}c({\bf k}) + {\bf q}\cdot ({\bf q}-{\bf k})c(|{\bf q}-{\bf k}|)\right]^2 .
\label{eq:vertMCT}
\end{eqnarray}
In a monoatomic fluid the direct correlation function $c(q)$  
is related to the static structure factor via the exact 
Ornstein-Zernike relation \cite{hansen} $\rho c(q)= 1-S^{-1}(q)$.
With all this, Eq. (\ref{eq:fMCT}) has been reduced 
to a closed set of coupled equations which can be solved self-consistently,
provided $S(q)$ and $c(q)$ are known (the latter are {\it external} inputs in the MCT equations).

For the case of systems with molecular architecture, Chong and Hirata
have obtained \cite{chonghirata}, by using projectors
over site-densities and site-currents, generalized MCT equations of motion for
site-site correlators $F_{ab}(q,t)$. The latter are defined as 
$F_{ab}(q,t) = n^{-1}\langle \rho_a(-q,0) \rho_b(q,t) \rangle $.
Note that $F_{ab}(q,0) = S_{ab}(q)$. The corresponding MCT equations of motion are
\begin{eqnarray}
\ddot{F}_{ab}(q,t)+\frac{q^2 k_{\rm B}T}{m}\sum_{x=1}^{N} S_{ax}^{-1}(q)F_{xb}(q,t) \nonumber \\
+\sum_{x=1}^{N}\int_{0}^{t}dt'M_{ax}(q,t-t')\dot{F}_{xb}(q,t')=0 ,
\label{eq:FabMCT}
\end{eqnarray}
where the memory kernel is now given by
\begin{eqnarray}
M_{ab}(q,t)  =  \frac{\rho_c k_{\rm B}T}{m q^2}\sum_{x,y=1}^{N}\int\frac{d^3 {\bf k}}{(2\pi)^3}\times \nonumber \\
 \{({{\bf q}}\cdot {\bf k})^2c_{ax}(k)c_{by}(k)F_{xy}(k,t)F_{ab}(p,t) \nonumber \\
+({{\bf q}}\cdot {\bf k})({{\bf q}}\cdot {\bf p})c_{ax}(k)c_{by}(p)F_{xb}(k,t)F_{ay}(p,t) \} ,
\label{eq:MabMCT}
\end{eqnarray}
with ${\bf p} = {\bf q} -{\bf k}$.
By comparing Eqs. (\ref{eq:FabMCT},\ref{eq:MabMCT}) 
with Eqs. (\ref{eq:fMCT},\ref{eq:mMCT},\ref{eq:vertMCT}) 
we note that the general mathematical structure 
of the kernel (bilinear in site-site correlators), and of the MCT equations of motion is retained.

Except for very small values of $N$, numerical solution of the MCT equations
for site-site correlators is extremely expensive, and further simplifications are needed
in order to obtain a tractable set of equations.
For the case of simple bead-spring chains, Chong and co-workers  have reduced \cite{chongprl,chongpre}
Eqs. (\ref{eq:FabMCT},\ref{eq:MabMCT}) to a scalar form for $f(q,t)$.
This is achieved by introducing in Eqs. (\ref{eq:FabMCT},\ref{eq:MabMCT})
the equivalent site, Eq. (\ref{eq:equivsite}), and ring, 
Eqs. (\ref{eq:ring},\ref{eq:invring}), approximations of the PRISM theory. 
The so-obtained scalar MCT equations of motion, memory kernel and vertex 
for polymer chains are formally {\it identical}
to Eqs. (\ref{eq:fMCT},\ref{eq:mMCT},\ref{eq:vertMCT}). 
The polymer character of the system only enters {\it implicitly} through
the PRISM relation $\rho c(q) =1/\omega(q) -1/S(q)$, which differs from the Ornstein-Zernike
equation, $\rho c(q)= 1-S^{-1}(q)$, for monoatomic systems. 
With this, general MCT predictions which originate from the mathematical structure of 
Eqs. (\ref{eq:fMCT},\ref{eq:mMCT},\ref{eq:vertMCT}) will be, due to the mentioned
formal equivalence, analogous both for monoatomic systems and for polymer chains. 
Now we summarize such general predictions.

In MCT, nonergodic arrested states (glasses) are defined as those for which
density correlators do not exhibit full relaxation. More specifically, 
if we introduce the nonergodicity parameters, defined as 
$f_q = \lim_{t \rightarrow \infty} f(q,t)$, MCT discriminates between
fluid states ($f_q = 0$) and glassy states ($f_q > 0$). 
At the MCT critical temperature $T_{\rm c}$, the nonergodicity parameters
jump from zero to nonzero values \cite{hopping}. In the following we use the notation
$f^{\rm c}_q$ for referring to the critical nonergodicity parameters,
i.e., the values of $f_q$ at $T = T_{\rm c}$.

By Laplace transform ($t \rightarrow z$) of Eqs. (\ref{eq:fMCT},\ref{eq:mMCT}) 
and taking the limit $z \rightarrow 0$, one finds a coupled set of equations
for the nonergodicity parameters:
\begin{equation}
\frac{f_q}{1 -f_q}= {\cal F}_q(\{f\})=
\int \frac{d^3 {\bf k}}{(2\pi)^3}{\cal V}({\bf q}-{\bf k},{\bf k})f_{|{\bf q}-{\bf k}|}f_k ,
\label{eq:nonergfac}
\end{equation}
where ${\cal F}_q(\{f\})$ denotes a functional, whose explict expression
is given in the right-hand side of the equation.
Note that Eq. (\ref{eq:nonergfac}) always has the trivial solution ${f_q =0}$.
Thus, glassy states take place when solutions $f_q >0$ also exist.

Given a tagged chain (labeled s), the density distribution for the $a$th monomer
of the tagged chain is defined as $\rho^{\rm s}_a (q)= \exp[i  {\bf q}\cdot {\bf r}^a_{\rm s}]$.
The site-site intrachain correlator is defined as 
$F^{\rm s}_{ab}(q,t) = \langle \rho^{\rm s}_a(-q,0) \rho^{\rm s}_b(q,t) \rangle$.
Note that  $F^{\rm s}_{ab}(q,0) = \omega_{ab}(q)$.
For the derivation of the MCT equations for $F_{ab}^{\rm s}(q,t)$ we refer to \cite{chongpre}. 
In this case the reduction to a scalar form is not possible. The corresponding
nonergodicity parameters $f_{ab}^s (q)=\lim_{t \rightarrow \infty} F_{ab}^s(q,t)$
are obtained by solving the $N\times N$-matrix equation \cite{chongdum}
\begin{equation}
f_{ab}^s(q)= \sum_{x,y=1}^N {\cal F}_{ax}^s(q)[{\bf I}+ {\boldmath \cal{ F}}_q^s]_{xy}^{-1} \omega_{yb}(q) , 
\label{eq:ssnonergpar}
\end{equation}
with ${\bf I}$ the identity matrix.
The corresponding functional ${\cal F}_{ab}^s (q)$ is given by
\begin{eqnarray}
{\cal F}_{ab}^s (q)=\sum_{x=1}^{N} \omega_{ax}(q)\int \frac{d^3 {\bf k}}{(2\pi)^3} \times \nonumber \\
{\cal V}^s({\bf q}-{\bf k},{\bf k})f_{xb}^s(k)f_{|{\bf q}-{\bf k}|} ,
\label{eq:ssMCTfunc}
\end{eqnarray}
with the vertex
\begin{equation}
{\cal V}^s({\bf q}-{\bf k},{\bf k})= \frac{\rho}{q^4}
S(|{\bf q}-{\bf k}|) [{\bf q}\cdot ({\bf q}-{\bf k})]^2 c^2(|{\bf q}-{\bf k}|) .
\label{eq:vertexselfab}
\end{equation}
The normalized self-correlator, usually introduced as 
$f^{\rm s} (q,t) = 
(nN)^{-1}\sum^{n}_{j=1}\sum^{N}_{a=1}\langle \exp[i {\bf q}\cdot ({\bf r}^a_j (t) - {\bf r}^a_j (0))]\rangle$,
can be equivalently obtained as  $f^{\rm s}(q,t) = N^{-1}\sum_{a=1}^{N}F_{aa}^s (q,t)$. 
Likewise, the corresponding nonergodicity parameters, defined as the long-time limit of  
$f^{\rm s}(q,t)$, can be obtained as $f_{q}^{\rm s}= N^{-1}\sum_{a=1}^{N}f_{aa}^s (q)$.
Thus, the solution of Eq. (\ref{eq:ssnonergpar}) also provides trivially the 
nonergodicity parameters for the self-correlator.

The separation parameter, $\epsilon = (T -T_{\rm c})/T_{\rm c}$, is introduced 
to quantify the relative distance to the critical temperature $T_{\rm c}$.
We are interested in the behavior of $f(q,t)$ in the ergodic fluid
by approaching $T_c$ from above. Thus we express the long-time behavior
of the density-density correlators as: 
\begin{equation}
f(q,t) = f^{\rm c}_q + g_q (t) ,
\label{eq:gqt}
\end{equation}
where $g_q (t)$ quantifies (small) deviations around $f^{\rm c}_q$ for $|\epsilon| \rightarrow 0$.
By introducing  Eq. (\ref{eq:gqt}) in Eqs. (\ref{eq:fMCT},\ref{eq:mMCT}), expanding
the functional ${\cal F}_q$ of Eq. (\ref{eq:nonergfac}) in a power series of $|\epsilon|$,
comparing the so-obtained resulting expressions and retaining the lower-order terms
(see, e.g., Ref. \cite{franosch} for a detailed exposition), one finds that $g_q (t) = h_q G(t)$,
where $h_q$ only depends on $q$, and $G(t)$ is a $q$-independent term which contains
the full time dependence of the deviations of $f(q,t)$ around $f^{\rm c}_q$.
Thus, we rewrite Eq. (\ref{eq:gqt}) as:
\begin{equation}
f(q,t) = f^{\rm c}_q + h_q G(t).
\label{eq:firstuniv}
\end{equation}

This expression is known as the {\it first universality} of the MCT
or {\it factorization theorem}. It predicts a scaling function $G(t)$ 
(known as the $\beta$-correlator)
that is {\it common} for all the density correlators (since it is $q$-independent).
Following the procedure mentioned in the previous paragraph \cite{franosch},
the function $G(t)$ is found to obey the equation:
\begin{equation}
\sigma -z^2 \tilde{G}^2 (z)  = \lambda z L[G^2 (t)]    ,
\label{eq:Gt}
\end{equation}
where $\tilde{G}(z)$ and $L[G^2 (t)]$ are the Laplace transform 
of respectively $G(t)$ and $G^2 (t)$. In this equation $\sigma =c|\epsilon|$,
with $c$ a constant (see \cite{franosch} for its explicit expression), and $\lambda$
is another constant given by
\begin{equation}
\lambda =   
\sum _{qk} e^{\rm T}_q C^{\rm c}(q,k,|{\bf q}-{\bf k}|)e_k e_{|{\bf q}-{\bf k}|}   .
\label{eq:lambda}
\end{equation}
The quantities $e_q$ and $e^{\rm T}_q$ are respectively 
the eigenvectors of the so-called stability matrix ${\bf C}^{\rm c}$
(see below) and its traspose,
with the normalization conditions $\sum_{q} e^{\rm T}_q e_q =1$
and $\sum_{q} e^{\rm T}_q (1-f^{\rm c}_q)e^2_q =1$. 
The elements of the stability matrix are given by
\begin{equation}
C^{\rm c}(q,k) = 
(1-f^{\rm c}_k)^2 \left(\frac{\partial {\cal F}_q}{\partial f_k}\right)_{\{f=f^{\rm c}\}} .
\label{eq:stabmat}
\end{equation}
The terms $C^{\rm c}(q,k,|{\bf q}-{\bf k}|)$ in Eq. (\ref{eq:lambda}) are given by:
\begin{eqnarray}
C^{\rm c}(q,k,|{\bf q}-{\bf k}|) = \hspace{4 cm} \nonumber \\
\frac{1}{2}(1-f^{\rm c}_k)^2 (1-f^{\rm c}_{|{\bf q}-{\bf k}|})^2
\left(\frac{\partial^2 {\cal F}_q}{\partial f_k \partial f_{|{\bf q}-{\bf k}|}}\right)_{\{f=f^{\rm c}\}} .
\label{eq:stabmat2}
\end{eqnarray}

Eq. (\ref{eq:Gt}) for the $\beta$-correlator does not have an analytical solution. Still,
asymptotic expressions can be obtained for different time windows. 
With this idea in mind the $\beta$-time scale is first  defined as
\begin{equation}
\tau_{\beta} = t_0 |\sigma|^{-1/(2a)}
\label{eq:taubeta}
\end{equation}
with $t_0$ a microscopic time scale and $a$
an exponent. The $\beta$-correlator is then rewritten as 
$G(t) = |\sigma|^{1/2} g_{\sigma}(t/\tau_{\beta})$. 
By introducing this expression in Eq. (\ref{eq:Gt})
and taking the limits $t \ll \tau_{\beta}$ and $t \gg \tau_{\beta}$ 
one finds the asymptotic solutions \cite{gotzebook}:
\begin{equation}
g_{\sigma}(t/\tau_{\beta}) = (t/\tau_{\beta})^{-a} \hspace{1cm} t \ll \tau_{\beta} ,
\label{eq:gpowa0}
\end{equation}
\begin{equation}
g_{\sigma}(t/\tau_{\beta}) = -B(t/\tau_{\beta})^b \hspace{1cm} t \gg \tau_{\beta} ,
\label{eq:gpowb0}
\end{equation}
where $B$ is a constant \cite{franosch}. The exponents $a$ and $b$ follow the constraint
\begin{equation}
\lambda= \frac{\Gamma^2 (1-a)}{\Gamma (1-2a)} = \frac{\Gamma^2 (1+b)}{\Gamma (1+2b)} ,
\label{eq:lambdacons}
\end{equation}
where $\Gamma$ denotes the Euler's Gamma function. 
According to Eqs. (\ref{eq:gpowa0},\ref{eq:gpowb0}), one finds
the asymptotic expressions for Eq. (\ref{eq:firstuniv}):
\begin{equation}
f(q,t) = f^{\rm c}_q + h_q (t/t_0)^{-a} \hspace{1cm} t \ll \tau_{\beta} ,
\label{eq:gpowa}
\end{equation}
\begin{equation}
f(q,t) = f^{\rm c}_q -h_q (t/\tau_{\alpha})^b \hspace{1cm} t \gg \tau_{\beta} .
\label{eq:gpowb}
\end{equation}
The analysis of the long-time decay usually includes higher-order corrections \cite{franosch}
to Eq. ({\ref{eq:gpowb}): 
\begin{equation}
f(q,t) = f^{\rm c}_q -h_q (t/\tau_{\alpha})^b +h_q^{(2)}(t/\tau_{\alpha})^{2b} +O(t^{3b}).
\label{eq:gpowb2}
\end{equation}
The latter is also known as the von Schweidler expansion.
In these equations $\tau_{\alpha}$ is the $\alpha$-time scale, defined as:
\begin{equation}
\tau_{\alpha} = B^{-1/b}t_0 |\sigma|^{-\gamma} .
\label{eq:taualphapow}
\end{equation}
The exponent $\gamma$ follows the constraint:
\begin{equation}
\gamma = \frac{1}{2a} + \frac{1}{2b} .
\label{eq:gamma}
\end{equation}

Another important prediction of the MCT for states approaching $T_{\rm c}$ from above, is the
{\it second universality} or {\it time-temperature superposition principle} (TTSP). 
This prediction arises as a long-time scaling property of the MCT equations of motion
\cite{gotzebook}. According to the TTSP, the long-time decay of any correlator $f(q,t)$ 
(i.e. the final part of the $\alpha$-relaxation) is invariant 
under scaling by the $\alpha$-relaxation time $\tau_{\alpha}$. 
In other words, for two temperatures $T_1$ and $T_2$ above $T_{\rm c}$ one finds
\begin{equation}
f(q,t/\tau_{\alpha}(T_1);T_1) = f(q,t/\tau_{\alpha}(T_2);T_2) =
\widetilde{f} (q,\hat{t}),
\label{eq:TTSP}
\end{equation}
where $\widetilde{f}(q,\hat{t})$ is a $T$-independent master function 
of the normalized time $\hat{t}$. While $G(t)$ is common to all correlators,
the master function $\widetilde{f}(q,\hat{t})$
associated to the TTSP is different for each correlator $f(q,t)$.
The superposition principle implies that the estimated
$\alpha$-relaxation time, defined in this work as the
time $\tau_{x}$ where  $f(q_{\rm max},t)$
takes a value $x$ well below the plateau, is proportional to
$\tau_{\alpha}$.  Thus, it also follows the asymptotic power law
\begin{equation}
 \tau_{x}(T) \propto (T-T_{\rm c})^{-\gamma} .
\label{eq:relaxtimepow}
\end{equation}

The $\alpha$-decay from the plateau to zero is often well described by an empirical
Kohlrausch-Williams-Watts (KWW) function,
\begin{equation}
f(q,t) = A_q\exp[-(t/\tau^{K}_q)^{\beta_q}] ,
\label{eq:KWW}
\end{equation}
with $A_q ,\beta_q < 1$.
Note that the latter does not come out as an analytical solution of the MCT equations. 
However in the  limit $q \rightarrow \infty$  of the KWW time 
$\tau^{K}_q$, MCT predicts that \cite{fuchsjcns}
\begin{equation}
\tau^{K}_q \propto q^{-1/b} \hspace{1 cm} q \rightarrow \infty, 
\label{eq:KWWtauq}
\end{equation}
where $b$ is the von Schweidler exponent introduced above.

The set of equations exposed in this section constitute a series
of universal results which originate from the structure of the MCT equations
of motion, Eqs. (\ref{eq:fMCT},\ref{eq:mMCT},\ref{eq:vertMCT}).
As mentioned above, the latter were initially derived
for simple hard-sphere systems, but the corresponding ones for polymer melts
become formally identical following the derivation by Chong and co-workers \cite{chongprl,chongpre}.
With this, the scaling laws exposed in this section 
will also hold in the MCT for polymer melts. Thus, the phenomenological 
analysis of our simulation data in terms of such scaling laws is justified 
within the framework of MCT. 
This analysis is presented in the next section.

\begin{center}
{\bf VI. MCT ANALYSIS OF SIMULATIONS}
\end{center}

In order to test the factorization theorem, Eq. (\ref{eq:firstuniv}), we compute the ratio:
\begin{equation}
R_{q}(t)=\frac{f(q,t)-f(q,t')}{f(q,t'')-f(q,t')}=
\frac{G(t)-G(t')}{G(t'')-G(t')}
\label{eq:factortest}
\end{equation}
where $t'$ and $t''$ are arbitrary times 
in the $\beta$-regime. The ratio for the self-correlators, $R^{\rm s}_{q}(t)$,
is defined analogously.
If the factorization theorem, and then also the right-hand side of Eq. (\ref{eq:factortest}),
is fulfilled, the ratios $R_{q}(t)$ and $R^{\rm s}_q(t)$ do not depend on the specific correlator.
Fig. \ref{fig:factheo} shows $R_{q}(t)$ and $R_{q}^{\rm s}(t)$ 
over a broad range of wave vectors  $2.3 \leq q \leq 16.5$. 
The data correspond to barrier strength $(K_{\rm B},K_{\rm T}) =(15,0.5)$
at $T=0.80$. The fixed times $t'' = 0.8$ and $t' = 100$ roughly correspond to the beginning 
and the end of the plateau regime. There is an intermediate time window of about
two decades where the data for density-density and self-correlators collapse onto 
a $q$-independent master curve, while they split at both short and late
times. Fig. \ref{fig:factheo2} demonstrates that the master curve is, 
moreover, the same for both density-density and self-correlators.
Thus, Figs. \ref{fig:factheo} and \ref{fig:factheo2} demonstrate 
the validity of the MCT first universality.

\begin{figure}
\begin{center}
\includegraphics[width=0.81\linewidth]{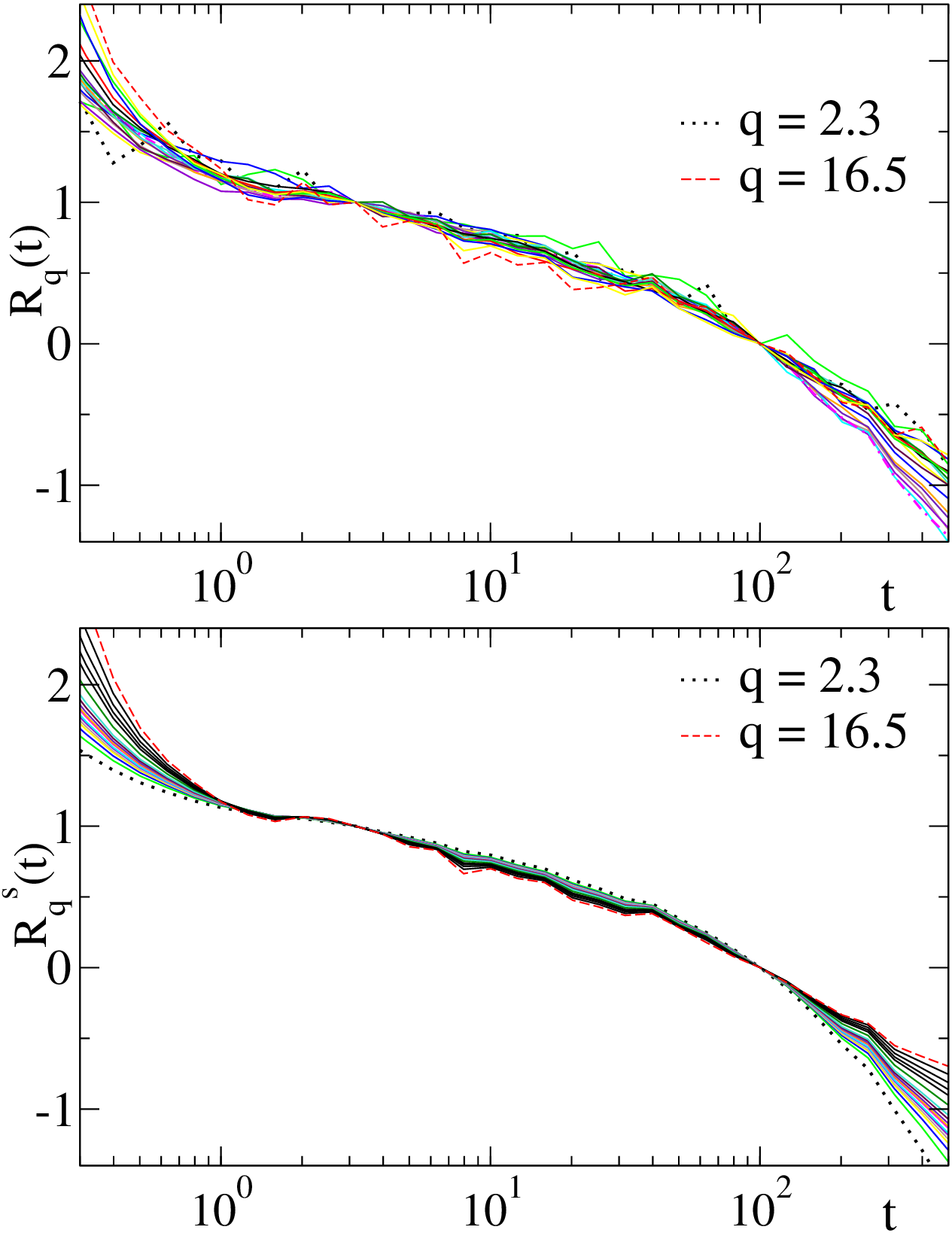}
\caption{Test of the factorization theorem, Eq. (\ref{eq:factortest}),
for density-density (top) and self-correlators (bottom)
at $T=0.80$ and $(K_{\rm B},K_{\rm T}) =(15,0.5)$. 
The different curves correspond to equispaced wave vectors in the range $2.3 \leq q \leq 16.5$.
The fixed times are $t'=100$ and $t'' = 0.8$.}
\label{fig:factheo}
\end{center}
\end{figure}
\begin{figure}
\begin{center}
\includegraphics[width=0.79\linewidth]{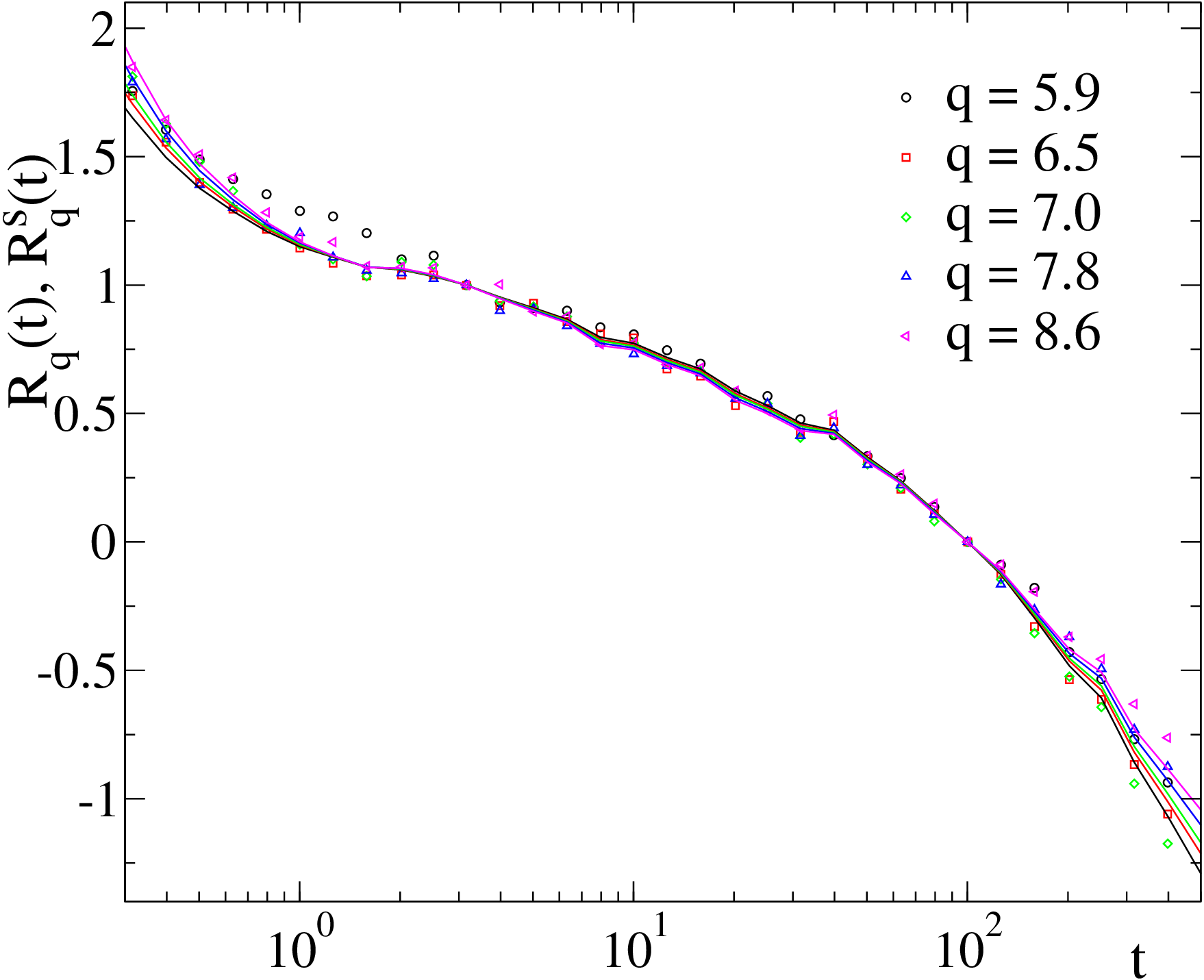}
\caption{Common representation of $R_{q}(t)$ (symbols) and $R_{q}^{\rm s}(t)$ (lines),
for selected wavevectors (common colors
correspond to common $q$-values).
As in Fig. \ref{fig:factheo}, data correspond to $T=0.80$ and $(K_{\rm B},K_{\rm T}) =(15,0.5)$,
and the selected fixed times are $t'=100$ and $t'' = 0.8$.} 
\label{fig:factheo2}
\end{center}
\end{figure}
\begin{figure}
\begin{center}
\includegraphics[width=0.80\linewidth]{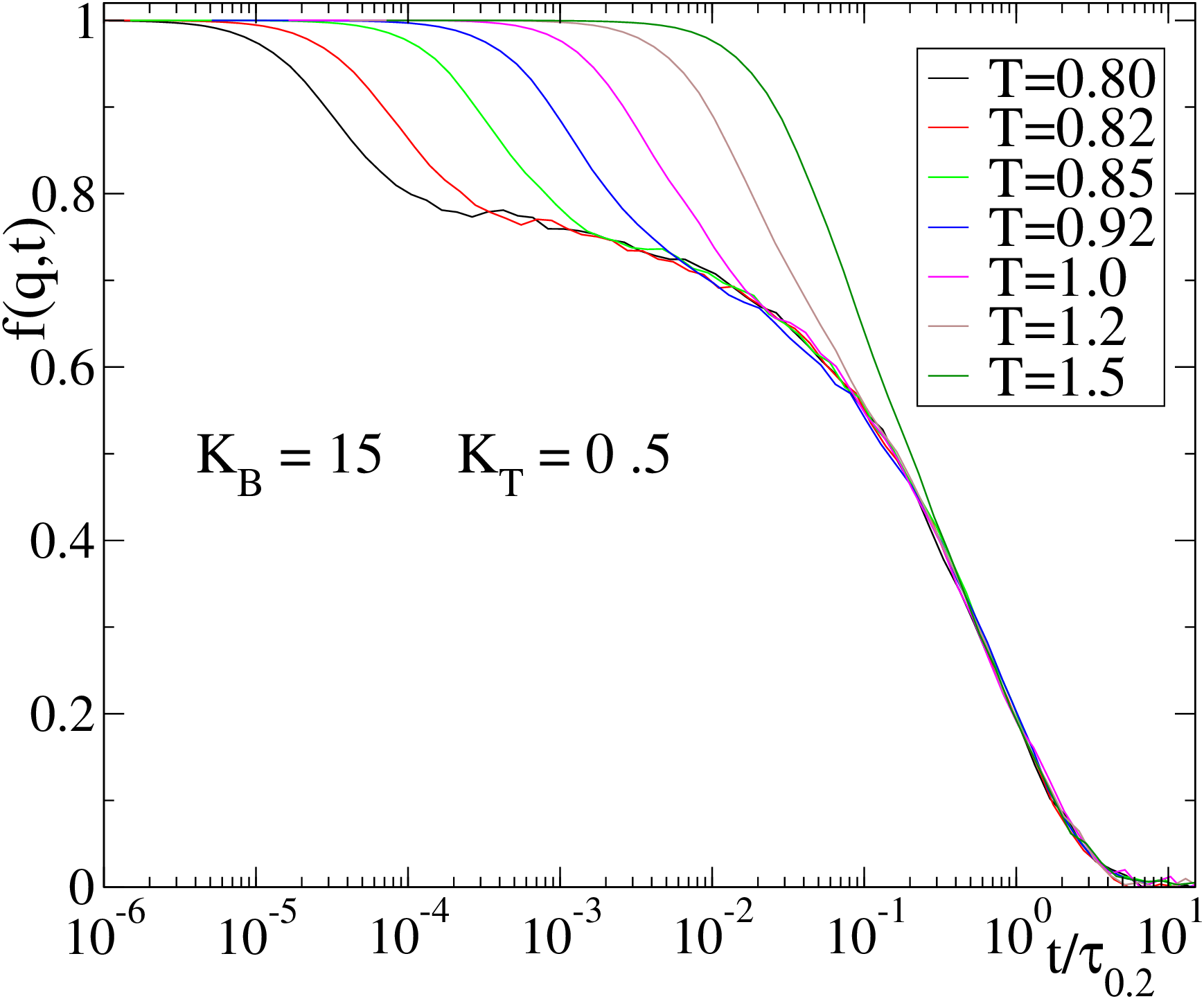}
\caption{Test of the time-temperature superposition principle, Eq. (\ref{eq:TTSP}),
for the density-density correlators 
at $q_{\rm max}\approx 7$, for the case  $(K_{\rm B}, K_{\rm T})= (15, 0.5)$.}
\label{fig:TTSP}
\end{center}
\end{figure}

Fig. \ref{fig:TTSP} shows a test of the TTSP, Eq. (\ref{eq:TTSP}), for the density-density correlator 
evaluated at  $q_{\rm max}$ (maximun of the static structure factor $S(q)$).
The data correspond to the case $(K_{\rm B},K_{\rm T}) =(15,0.5)$
and cover a broad temperature range $0.80 \leq T \leq 1.5$.
Data collapse onto a master curve after rescaling the absolute time by the
relaxation time $\tau_{0.2}$.
Thus, the MCT second universality also holds for chains with strong intramolecular barriers.

\begin{figure}
\begin{center}
\includegraphics[width=0.9\linewidth]{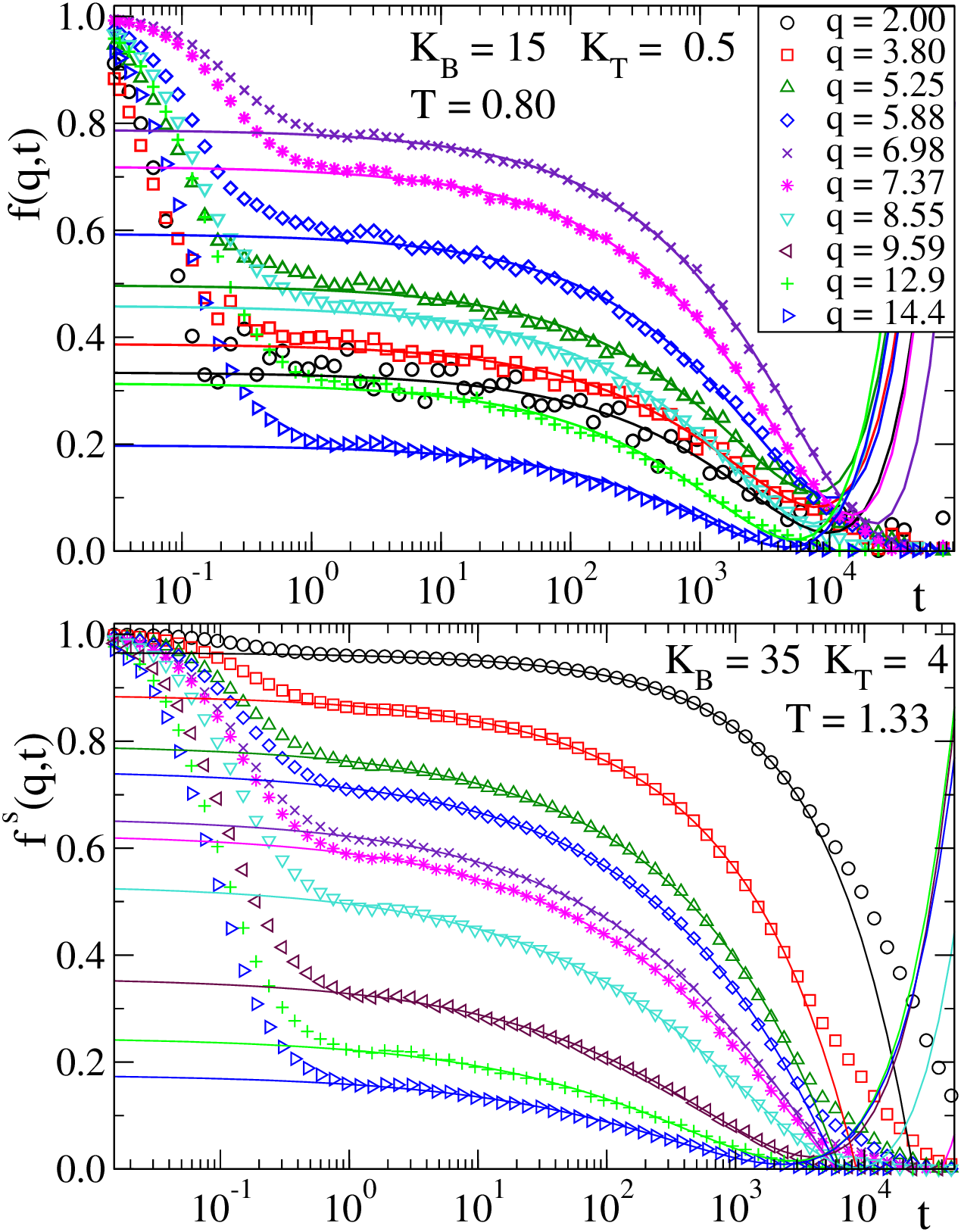}
\caption{Symbols: simulations results  for density correlators. 
Top panel: $f(q,t)$ for $(K_{\rm B} ,K_{\rm T}) = (15, 0.5)$, at $T = 0.80$. 
Bottom panel: $f^{\rm s}(q,t)$ for $(K_{\rm B}, K_{\rm T} )= (35, 4)$, at $T = 1.33$. 
Identical symbols in both panels correspond to identical wave vectors $q$
[values are given in panel (a)]. Lines are fits to the von Schweidler expansion,
Eq. (\ref{eq:gpowb2}) (up to second-order terms), with $b = 0.50$ (top) and 0.37 (bottom).} 
\label{fig:vonsch}
\end{center}
\end{figure}

Solving numerically the MCT equations and determining the dynamic exponents $(a,b,\gamma,\lambda)$ 
is in general a difficult task. When numerical solutions are not available, nonergodicity parameters, 
prefactors and exponents in Eqs. (\ref{eq:gpowa},\ref{eq:gpowb2},\ref{eq:relaxtimepow},\ref{eq:KWWtauq})
can be obtained as fit parameters from simulation 
or experimental data (see, e.g., Refs. \cite{mctrev2,aichele,vanmegen,koband2,baschrev}). 
Consistency of the analysis requires that dynamic correlators and relaxation times 
are described by a common set of exponents, all of them related to a single
$\lambda$-parameter through Eqs. (\ref{eq:lambdacons},\ref{eq:gamma}).

We have performed this consistency test for 
all the investigated range of barrier strength.
The following figures in this section illustrate, for some representative cases,
the analysis of simulation data in terms of MCT asymptotic laws.
Fig. \ref{fig:vonsch} shows for a broad $q$-range ($2.0 \leq q \leq 14.4$), 
fits to the von Schweidler expansion, Eq. (\ref{eq:gpowb2}) (up to second-order terms).
Data correspond to density-density correlators
$f(q,t)$ for the state point $(K_{\rm B} ,K_{\rm T}) = (15, 0.5)$, $T = 0.80$ (labelled S1), 
and to self-correlators $f^{\rm s}(q,t)$ for 
$(K_{\rm B}, K_{\rm T} )= (35, 4)$, $T = 1.33$ (labelled S2). 
A good description of the simulation data
is achieved, for all the range of $q$-values and over almost four time decades, 
with a fixed $b$-exponent ($b = 0.50$ and 0.37 for respectively S1 and S2).

Fig. \ref{fig:fne} displays, for the former values of the barrier strength, 
the $q$-dependence of the so-obtained critical nonergodicity 
parameters ($f_q^{\rm c}$ for $f(q,t)$  and $f_q^{\rm sc}$ for $f^{\rm s}(q,t)$). 
For comparison, we also include the fully-flexible case 
$(K_{\rm B} ,K_{\rm T}) = (0, 0)$. As deduced from the stronger decay of 
$f_q^{\rm c}$ and $f_q^{\rm sc}$ for stronger barriers, the introduction
of chain stiffness yields a weaker stability of density fluctuations. 
It also induces a weaker localization for self-motions at fixed density. 
Thus, by making an approximate fit of $f_q^{\rm sc}$ to Gaussian behavior, 
$f_q^{\rm sc} \approx \exp(-q^2 l_{\rm c}^2 /6)$, we estimate, 
at fixed $\rho =1.0$, a localization length $l_{\rm c} =$ 0.19, 0.21, and 0.23 for respectively  
$(K_{\rm B},K_{\rm T}) =$ (0,0), (15,0.5), and (35,4).

\begin{figure}
\begin{center}
\includegraphics[width=0.85\linewidth]{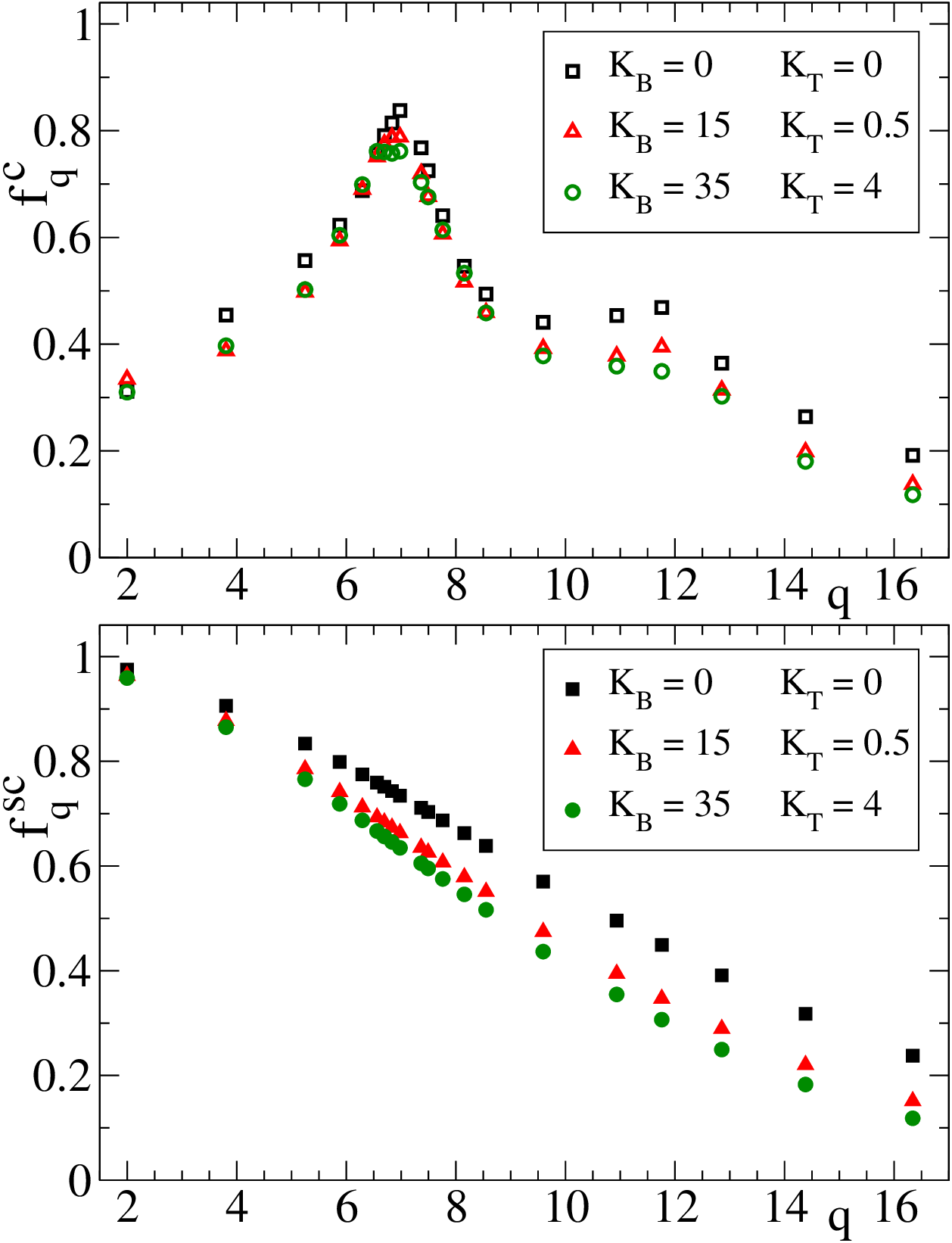}
\caption{Critical nonergodicity parameters, as determined
from fits to Eq. (\ref{eq:gpowb2}), for different barrier strength.
Top and bottom panels show data for respectively $f_q^{\rm c}$
and $f_q^{\rm sc}$.} 
\label{fig:fne}
\end{center}
\end{figure}

Data of self-correlators from the plateau to the limit of the simulation window have
been fitted to KWW functions, Eq. (\ref{eq:KWW}) (not shown). Fig. \ref{fig:tauq} shows the $q$-dependence of the 
so-obtained KWW relaxation times $\tau_q^{\rm K}$ for the former values of the barrier strength,
at their respective lowest investigated temperatures.
The lines represent tests of the MCT prediction $\tau_q^{\rm K} \propto q^{-1/b}$ for large $q$. 
A good description of the data is obtained with the same $b$-exponents used for the independently
obtained von Schweidler fits of Fig. \ref{fig:vonsch}.


\begin{figure}
\begin{center}
\includegraphics[width=0.65\linewidth]{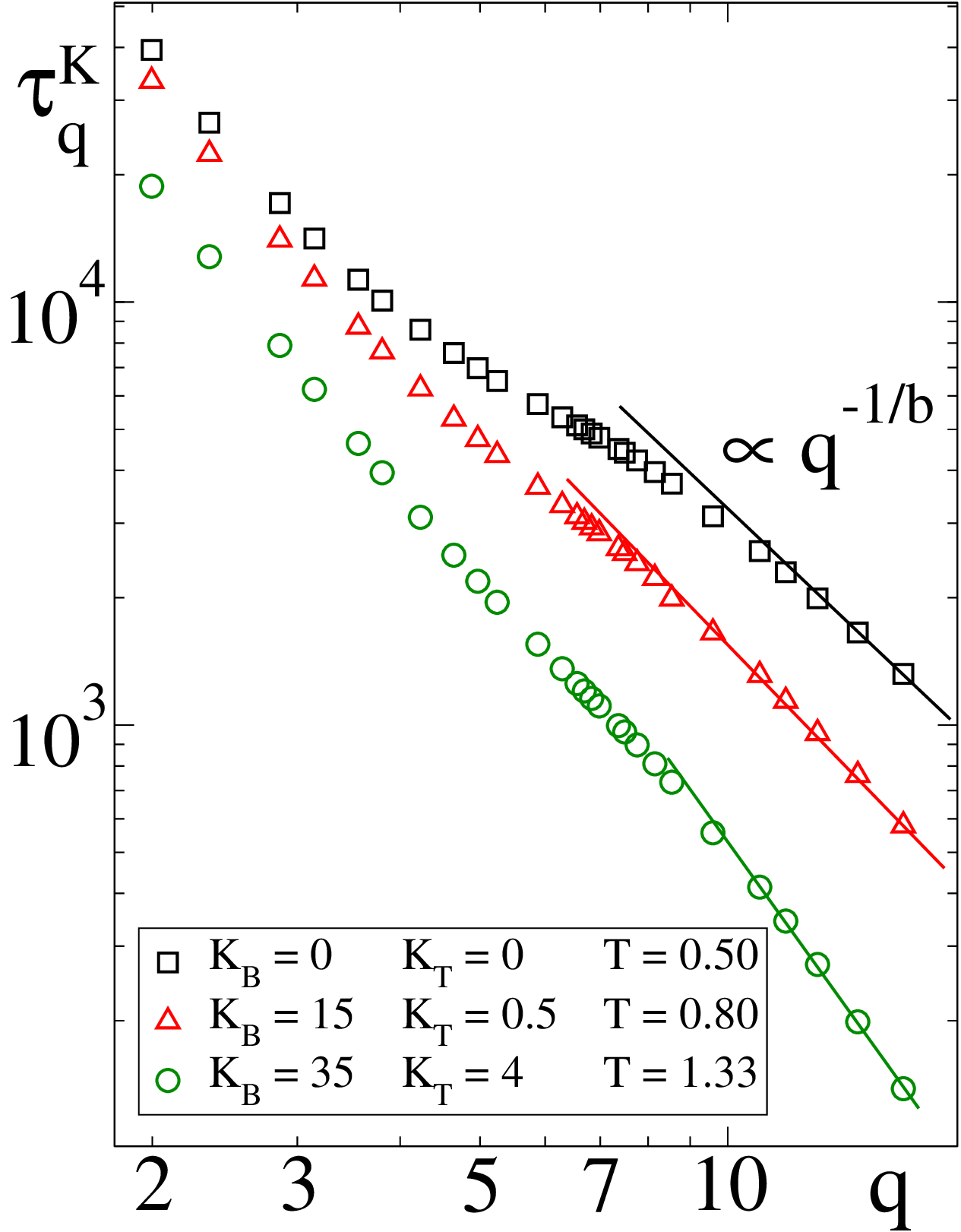}
\caption{Symbols: $q$-dependence of KWW relaxation times 
for different barrier strength. Lines are fits to $\propto q^{-1/b}$  (see text).
From top to bottom $b = 0.54$, 0.50 and 0.37.} 
\label{fig:tauq}
\end{center}
\end{figure}

\begin{figure}
\begin{center}
\includegraphics[width=0.75\linewidth]{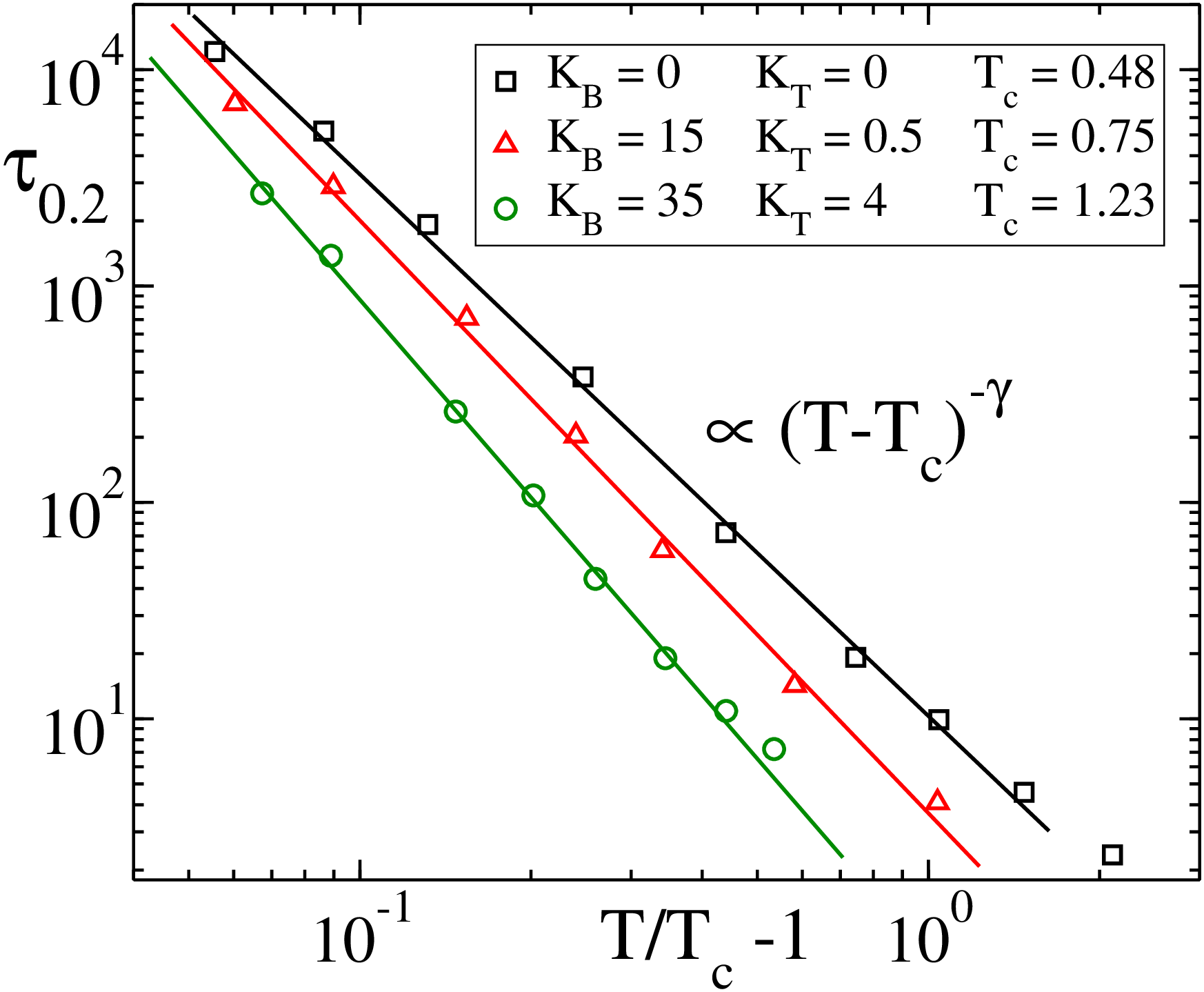}
\caption{Symbols: $T$-dependence of relaxation times $\tau_{0.2}$ 
for different barrier strength. Lines are fits
to   $\propto (T-T_{\rm c})^{-\gamma}$ (see text).
From top to bottom $\gamma = 2.60$, 2.74 and 3.43.} 
\label{fig:tauTc}
\end{center}
\end{figure}

Fig. \ref{fig:tauTc} shows, for the same values of $(K_{\rm B} ,K_{\rm T})$ in Fig. \ref{fig:tauq},
a test of the power law $\tau_{0.2} \propto (T-T_{\rm c})^{-\gamma}$
for the temperature dependence of the estimated $\alpha$-relaxation times. 
The fit covers about three time decades. By representing the data
in terms of the separation parameter $T/T_{\rm c} -1$, clearly different $\gamma$-exponents
are evidenced for different barrier strength. A good description of the data is obtained
with the $\gamma$-values derived, through Eqs. (\ref{eq:lambdacons}, \ref{eq:gamma}), 
from the $b$-values used in Figs. \ref{fig:vonsch} and \ref{fig:tauq}. 
This result demonstrates the consistency of the MCT analysis for the representative
examples showed here, which cover all the range of investigated barrier strength
between fully-flexible and stiff chains.

\begin{table}
\begin{center}
\begin{tabular}{cccccccccccc}
&$\rho$ &\ &$K_{\rm B}$  &\ &$K_{\rm T}$  &\ &$R_{\rm ee}^{\rm c}$  &\ &$T_{\rm c}$  &\ &$\lambda$ \\
\hline
\\
&1    &\ &0         &\ &0                 &\ &3.6             &\ &0.48          &\ &0.761 \\
&1    &\ &4         &\ &0.1               &\ &4.4             &\ &0.54          &\ &0.767 \\
&1    &\ &8         &\ &0.2               &\ &4.7             &\ &0.67          &\ &0.773 \\
&1    &\ &15        &\ &0.5               &\ &5.2             &\ &0.75          &\ &0.785 \\
&1    &\ &25        &\ &1                 &\ &5.5             &\ &0.82          &\ &0.827 \\
&1    &\ &25        &\ &4                 &\ &6.4             &\ &1.02          &\ &0.845 \\ 
&1    &\ &35        &\ &4                 &\ &6.5             &\ &1.23          &\ &0.862 \\
&0.93 &\ &35        &\ &4                 &\ &6.9             &\ &1.02          &\ &0.885 \\  
\\
\hline
\end{tabular}
\end{center}
\caption{Values of the MCT $\lambda$-exponents and critical temperatures $T_{\rm c}$
for different $\rho$ and barrier strength. Also included
are the mean chain end-to-end radii $R_{\rm ee}^{\rm c}$ at $T_{\rm c}$.}
\label{tab:paramkbkt}
\end{table}

Similar consistent tests (not shown) have been performed for the rest of investigated systems.
Table \ref{tab:paramkbkt} displays the results for  the so-obtained $\lambda$-exponents and 
critical temperatures $T_{c}$ as a function of $(K_{\rm B} ,K_{\rm T})$. We also include the corresponding
value of the mean end-to-end radius (computed at $T_{\rm c}$), which provides a qualitative 
characterization of chain stiffness. From the numerical values in Table \ref{tab:paramkbkt} 
a clear correlation between the strength 
of the internal barriers and the values of $T_{\rm c}$ and $\lambda$ is unambiguously demonstrated. 
The interplay between packing effects and
intramolecular barriers induces a progressive increase of $T_{\rm c}$ at fixed density.
A similar effect is observed for the $\lambda$-exponent, which increases from 
$\lambda = 0.761$ for fully-flexible chains to $\lambda = 0.885$ for the stiffest investigated chains.
The smallest $\lambda$-values in Table \ref{tab:paramkbkt} 
are typical of simple glass-formers as the archetype hard-sphere fluid
($\lambda = 0.74$) \cite{franosch} where dynamic arrest is driven by packing effects. 
The largest ones, $\lambda \lesssim 0.9$,
are similar to those observed in realistic models of polymer melts which 
incorporate the chemical structure of the chains. Some examples include 
poly(vinyl methylether) \cite{capponi}, polybutadiene \cite{narros1} 
or poly(vinyl ethylene) \cite{narros2}, with respective values of $\lambda =$  0.87, 0.93, and 0.93.

Thus, the analysis presented here rationalizes the difference in the MCT exponents 
between fully-flexible bead-spring models and real polymers. 
The systematic study performed by tuning the barrier strength
suggests that large $\lambda$-exponents in real polymers arise
from the interplay between two distinct mechanisms for dynamic arrest. These are
general packing effects and polymer-specific intramolecular barriers. Large $\lambda$-values
arising from the interplay between distinct arrest mechanisms have been observed in systems
of very different nature, as short-ranged attractive 
colloids \cite{sperl,dawson,zaccarelli} (competition between hard-sphere repulsion
and short-ranged reversible bonding), polymer blends \cite{blends1,blends2} and colloidal mixtures with strong dynamic 
asymmetry \cite{mixtures1,mixtures2} (bulk-like caging and matrix-induced confinement), 
or densified silica \cite{voigtsildens} (presumably bonding and packing). 
Numerical solutions of the MCT equations in
short-ranged attractive colloids \cite{sperl,dawson}
and quenched-annealed mixtures \cite{krakoviack} have revealed the existence
of higher-order MCT transitions, which are characterized by the upper limit $\lambda =1$.
Whether higher-order MCT transitions are present at some region of the control parameter space
of the investigated model is an open question.

In this section we have performed a phenomenological analysis
of the simulation data within the framework of MCT. 
In the next section the observed trends are compared with numerical solutions of the MCT equations.

\begin{center}
{\bf VII. SOLUTION OF THE MCT EQUATIONS}
\end{center}

We have solved Eqs. (\ref{eq:nonergfac}) and (\ref{eq:ssnonergpar}) for 
the nonergodicity parameters, for all the investigated range of barrier strength.
In analogy with the procedure exposed in, e.g., Refs. \cite{franosch,fuchspre},
the integrals over the reciprocal space in the corresponding MCT functionals of 
Eqs. (\ref{eq:nonergfac}}) and (\ref{eq:ssMCTfunc}) are discretised to a grid
of $M = 600$ equispaced points, with $q$-spacing $\Delta q = 0.1$, leading to the
expressions:
\begin{eqnarray}
\frac{f_q}{1-f_q} = \frac{\rho (\Delta q)^3}{32 \pi^2} \sum_{x_k}\sum_{x_p}'\frac{x_k x_p}{x_q^5}S(q)S(k)S(p)   \nonumber \\
\times [(x_q^2+x_k^2-x_p^2)c(k)+(x_q^2+x_p^2-x_k^2)c(p)]^2f_k f_p
\label{eq:nonergfacdis}
\end{eqnarray}
and
\begin{eqnarray}
{\cal F}_{ab}^s (q)= \frac{\rho (\Delta q)^3}{16 \pi^2}\sum_{x=1}^{N} \omega_{ax}(q) \times \nonumber \\
\sum_{x_k}\sum_{x_p}' 
\frac{x_k x_p}{x_q^5}S(p)[(x_q^2+x_p^2-x_k^2) c(p)]^2 f_{xb}^s(k) f_p .
\label{eq:ssMCTfuncdisc}
\end{eqnarray}
In these expressions the wavevectors are defined as 
$q = x_q \Delta q$, $k = x_k \Delta q$, and $p = x_p \Delta q$, 
with $x_q,x_k,x_p=1/2,3/2,...1199/2$. The prime at the sums over $x_p$ means that 
the latter are restricted
to $x_p$-values following the condition $|x_q-x_k|+1/2 \leq x_p \leq  x_q+x_k-1/2$.

The solutions of Eq. (\ref{eq:nonergfacdis}) are found by a standard iterative procedure 
$f_{q}^{j+1} /[1-f_{q}^{j+1}]={\cal F}_q(\{f^{j}\})$, with $j$ the iteration 
step, and with the initial condition $f_{q}^{0}=1$. 
It can be demonstrated that the stability matrix in Eq. (\ref{eq:stabmat})
has always a maximum non-degenerate eigenvalue
$E \le 1$, which takes the upper value $E_{\rm c} =1$ at the critical point \cite{franosch}. 
Thus, by following the drift of $E$ with changing temperature it is possible to bracket
the values of the critical nonergodicity parameters $f^{\rm c}_q$,
and the critical temperature $T_{\rm c}$, with very high precision.
Once the values of $f^{\rm c}_q$ are obtained, they are fixed in the functional 
of Eq. (\ref{eq:ssMCTfuncdisc}), and a small number of iterations
is needed to find the corresponding critical values $f^{\rm sc}_{ab}(q)$.
Finally, the critical nonergodicity parameters for self-correlations are obtained
as $f^{\rm sc}_q = N^{-1}\sum_{a=1}^{N} f^{\rm sc}_{aa}(q)$.

Following the procedure exposed above, we solved Eq. (\ref{eq:nonergfacdis}) by inserting as external inputs
the structural quantities, $S(q)$ and $c(q)$, as directly computed from the simulations.
However, as previously reported in Ref. \cite{chongpre} for fully-flexible chains,
a MCT transition was not observed for any of the investigated barrier strength. 
This means that the theoretical critical temperature $T_{\rm c}$ is below the lowest simulation temperature 
for which equilibration was possible. This result is different from the usual observation in
non-polymeric systems, for which the theoretical critical point is accessible in simulation time scales.
The reason of this difference is, in some way, related with the unability to crystallize 
of bead-spring models, which avoids a fast growing of peaks under cooling in 
the static structure factor $S(q)$, leading to MCT kernels which are not sufficiently strong
to provide nonzero solutions of $f_q$.

\begin{figure}
\begin{center}
\includegraphics[width=0.85\linewidth]{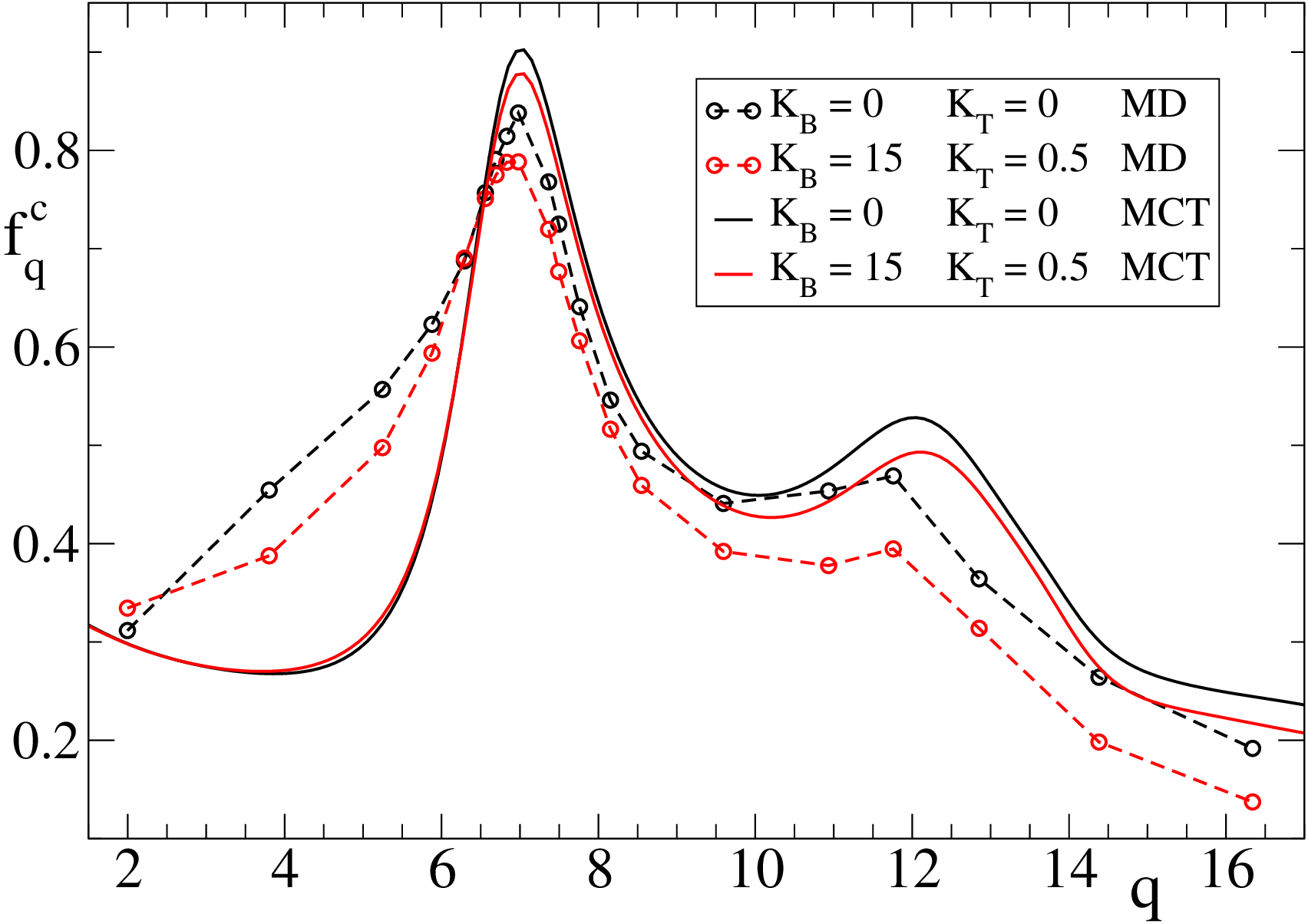} 
\newline
\newline
\includegraphics[width=0.85\linewidth]{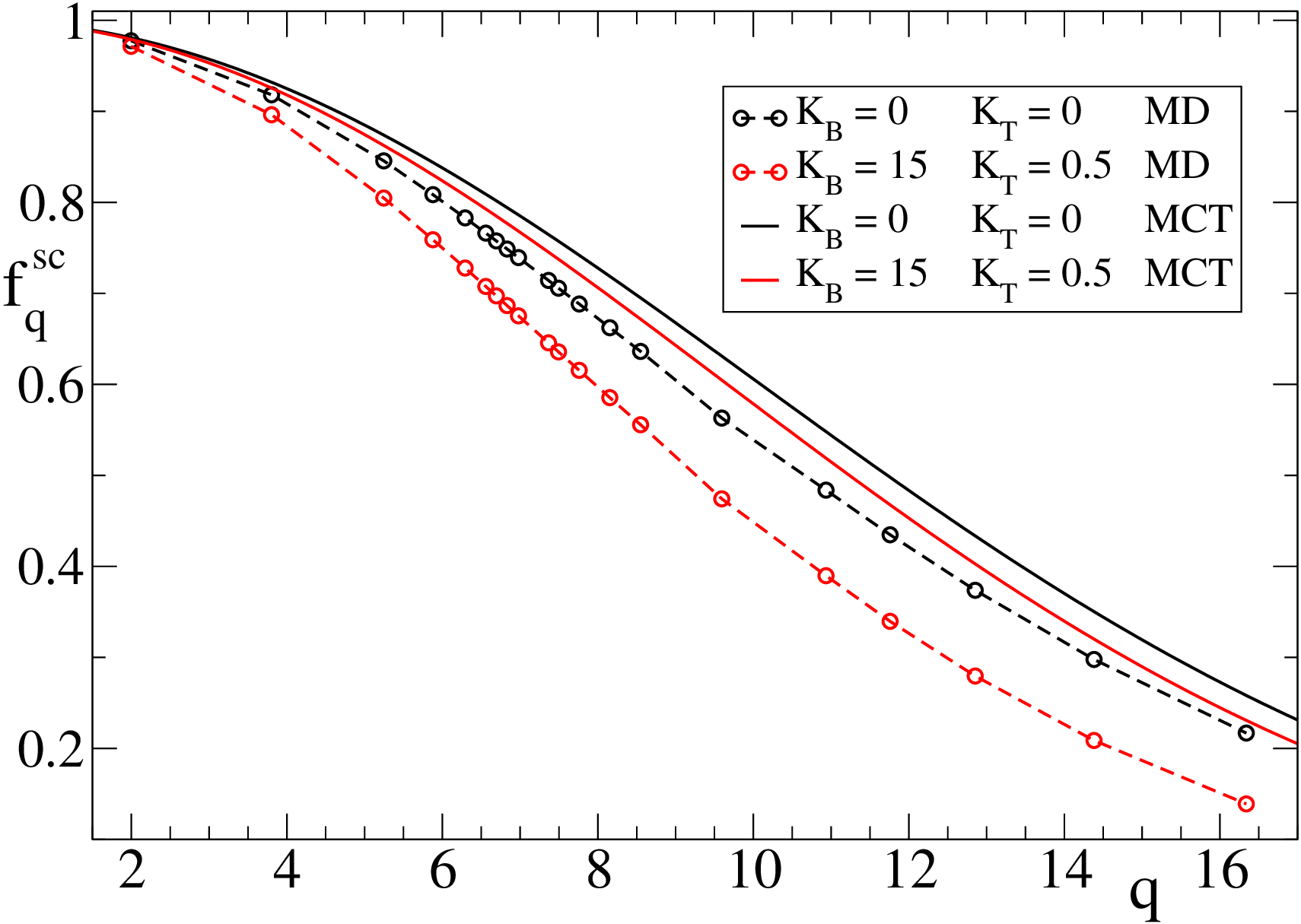} 
\newline
\caption{Comparison of the critical nonergodicity parameters 
$f_q^{\rm c}$ (top) and $f_q^{\rm sc}$ (bottom)
obtained from MCT calculations (solid lines) with those obtained
from the phenomenological analysis of the simulation data (symbols, dashed lines are guides for the eyes). 
Results are shown both for fully-flexible chains
and for representative stiff chains with $(K_{\rm B},K_{\rm T}) =(15,0.5)$.}
\label{fig:nonergpar}
\end{center}
\end{figure}

\begin{figure}
\begin{center}
\includegraphics[width=0.7\linewidth]{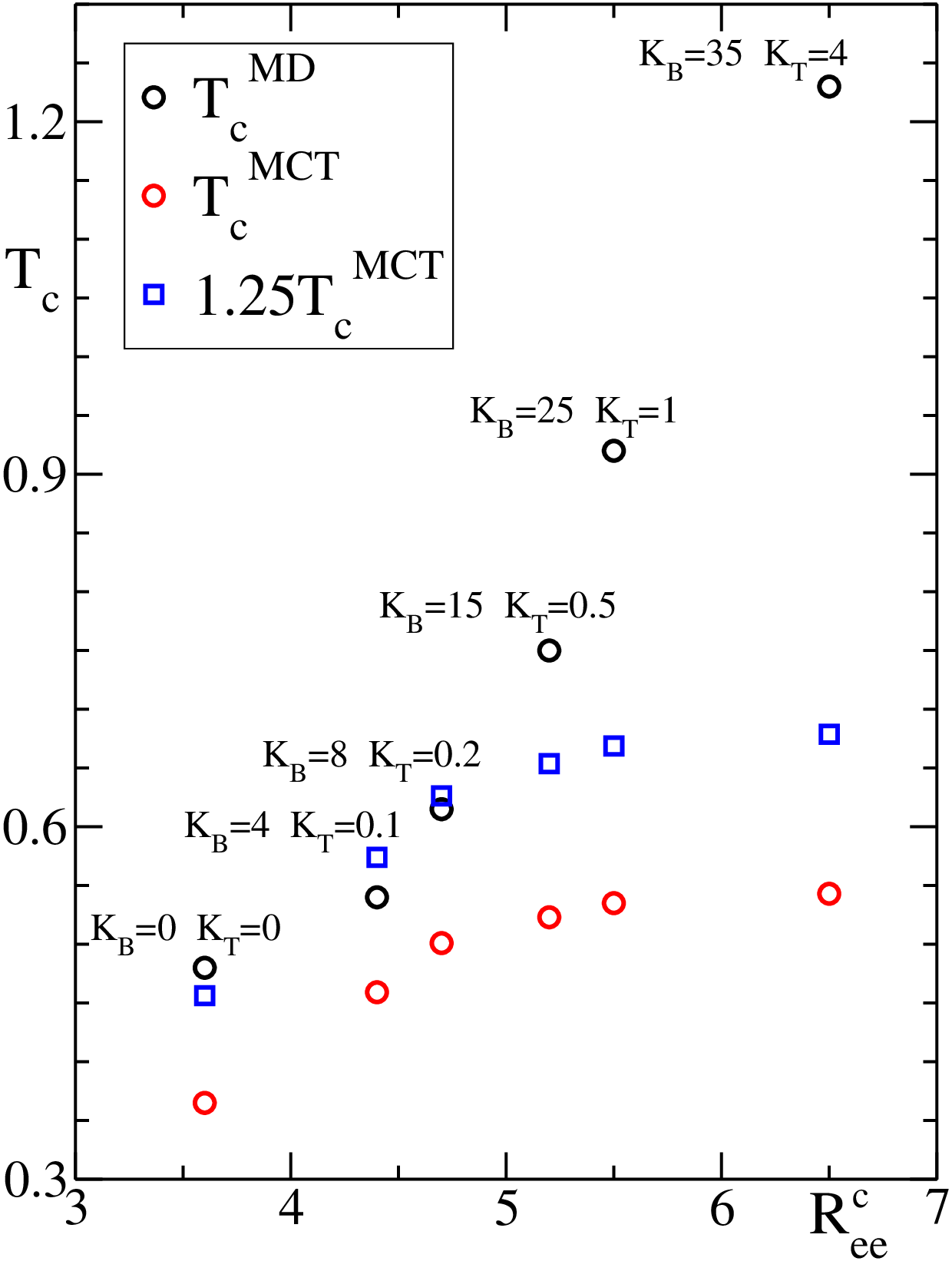} 
\caption{Critical temperature $T_{\rm c}$ as a function of the end-to-end radius $R_{\rm ee}^{\rm c}$.  
The theoretical $T_{\rm c}^{\rm MCT}$, as obtained by solving the MCT equations, 
is compared with the $T_{\rm c}^{\rm MD}$ from the phenomenological analysis of the simulation data.}
\label{fig:Tc}
\end{center}
\end{figure}

Since static correlations computed from our equilibrium simulations do not induce a MCT transition, we are forced 
to use a structural theory for estimating $S(q)$ and $c(q)$ at lower temperatures,
which will allows us to insert them in the MCT equations and to search for the critical temperature. 
Thus, we solve numerically the PRISM equation
\begin{equation}
\rho c(q)= 1/\omega(q)-1/S(q) ,
\label{eq:prism2}
\end{equation}
with the Percus-Yevick (PY) closure relation \cite{hansen} for the 
non-bonded potential $V(r)$ of Eq. ({\ref{eq:potsoft}}). The PY relation is given by:
\begin{equation}
c(r) = [1-\exp(V(r)/k_B T)](h(r)+1) ,
\label{eq:PY}
\end{equation}
where $c(r)$ and $h(r)$ are the Fourier transforms in the real space
of $c(q)$ and $h(q)$.
The coupled set of nonlinear equations (\ref{eq:prism2},\ref{eq:PY}) is solved by a
standard Picard iteration method \cite{notegridpicard}
for the quantity $\Gamma (r) = h(r)-c(r)$, 
which is a smooth function over all the range of $r$. 
The form factor $\omega(q)$ is an {\it external} input in this procedure.
We observed  (see above) that $\omega(q)$ exhibits a very weak temperature dependence
in comparison to the total static structure factor $S(q)$.
Thus we just use for each barrier strength the $\omega(q)$, 
as computed from the simulations, at the lowest temperature for which equilibration was possible.

In Fig. \ref{fig:nonergpar} we show a comparison
of  the critical nonergodicity parameters $f^{\rm c}_q$ (top panel) 
and $f^{\rm sc}_q$ (bottom panel)
as obtained from numerical solution of the MCT equations,
with the results of the fitting procedure of simulation data (see above).
The theoretical results qualitatively reproduce the simulation trends,
and in particular the observation that at fixed density the intramolecular barriers 
induce a weaker localization length. Quantitatively, the MCT solutions overstimate
the amplitude of the nonergodicity parameters, except in the low-$q$ 
region of $f_q^{\rm c}$, for which MCT clearly understimates the results.

In figure \ref{fig:Tc} we show a representation of  the critical temperature $T_{\rm c}$
as a function of the end-to-end radius $R_{\rm ee}^{\rm c}$, 
which quantifies chain stiffness. Values of $T_{\rm c}$ obtained from the 
phenomenological analysis of the simulations ($T_{\rm c}^{\rm MD}$)  
and from the numerical solutions of the MCT equations ($T_{\rm c}^{\rm MCT}$) are compared.
We note that $T_{\rm c}^{\rm MD}$ seems to grow monotonously 
with chain stiffness.  This  trend is well reproduced by
the theory for low and moderate values of the internal barriers. 
Thus, for values of  bending and torsional constants $K_B<15$ and $K_T<0.5$, 
the dependence of $T_{\rm c}^{\rm MCT}$ on $R_{\rm ee}^{\rm c}$  roughly displays the same slope
as for $T_{\rm c}^{\rm MD}$, with a shift factor $T_{\rm c}^{\rm MD}/T_{\rm c}^{\rm MCT} \approx 1.25$. 
Similar shifts between simulation and theory, which have their origin 
in the mean-field character of the MCT, are observed in other systems \cite{vanmegen,puertas,sciortino}.
The range of barrier strength for which $T_{\rm c}^{\rm MCT}$ and $T_{\rm c}^{\rm MD}$ 
are roughly parallel is significant. Note that for $(K_{\rm B} ,K_{\rm T}) = (8,0.2)$
the end-to-end radius $R_{\rm ee}^{\rm c}$ is a factor 1.3 longer than for fully-flexible chains.
 
By further increasing chain stiffness the differences between $T_{\rm c}^{\rm MD}$ 
and $T_{\rm c}^{\rm MCT}$ progressively increase. 
We observe a saturation of the theoretical $T_{\rm c}^{\rm MCT}$ around $\approx 0.55$,
while the simulation $T_{\rm c}^{\rm MD}$ grows up to a value of 1.23 for the
stiffest investigated chains. Thus, the agreement between theory and simulation
clearly breaks  for stiff chains.

Finally, we have computed the corresponding theoretical $\lambda$-exponents 
according to the definitions of Eqs. (\ref{eq:lambda},\ref{eq:stabmat},\ref{eq:stabmat2}). 
We find an almost constant value of $\lambda \approx 0.72$
for all the investigated range of barrier strength.
This result is clearly different from the observations in the phenomenological MCT analysis
of simulation data (see data in Table \ref{tab:paramkbkt}), which provides 
a strong dependence of $\lambda$ on the barrier strength.

\begin{center}
{\bf VIII. DISCUSSION}
\end{center}

The results reported in the previous section show that, 
though reproducing some qualitative simulation trends for low and moderate barriers,
numerical solutions of the MCT equations exhibit important differences
with simulation values as the limit of stiff chains is approached.
Another result to be understood is the clear disagreement between the almost
constant value of the theoretical $\lambda$-exponent,
and the observed strong dependence of simulation values on the barrier strength. 

The observed disagreement between theory and simulation for strong barriers does not seem
to be related with the failure of the PRISM approximations for stiff chains, which
have been introduced in the derivation of the MCT equations. Indeed we have shown that the quality
of the used PRISM approximations is the same for fully-flexible and stiff chains
(Figs. \ref{fig:eqsite}, \ref{fig:eqsite2} and \ref{fig:ring}).
Having said this, it might be argued that the theory is simply wrong: the 
phenomenological MCT analysis is apparently successful, but one finds that it has little to do
with the theory, for stiff chains, when solving the MCT equations.
However, we remind that the phenomenological analysis has shown, 
for all the investigated range of barrier strength:
i) the validity of the two MCT universalities, i.e., 
the factorization theorem (Figs. \ref{fig:factheo} and \ref{fig:factheo2}) 
and the TTSP (Fig. \ref{fig:TTSP}) ii) the possibility of a good description of
different dynamic observables (Figs. \ref{fig:vonsch}, \ref{fig:tauq} and \ref{fig:tauTc}) 
with a set of dynamic exponents which are consistently transformed,
through Eqs. (\ref{eq:lambdacons},\ref{eq:gamma}), to a single $\lambda$-exponent. 

We believe that all these observations, for all the investigated cases, are not fortuitous.
At this point it must be noted that the predictions referred to in points i) and ii)
arise, within MCT, as a consequence of the mathematical structure of the equations of motion, 
more precisely they originate from the bilinear form of the memory kernel. The specific values of the numerical solutions
clearly depend on the coefficients of the bilinear products (which enter through the vertices of the kernel),
but the factorization theorem, the TTSP, and the asymptotic scaling laws are universal properties
provided the kernel is bilinear. Thus, the results
of the phenomenological analysis suggest that the underlying physics may be connected
to a bilinear memory kernel, though for high barriers the actual coefficients strongly
differ from  those introduced by MCT through the vertices, thus leading to theoretical results which
strongly differ from simulations.

In other words, the present results suggest
that there may be relevant static contributions for the case of stiff chains
which are missing in the MCT vertices.  Thus, the inclusion of such
contributions will increase the strength
of the kernel and will induce the theoretical transition at higher values of $T_{\rm c}$,
which might improve the comparison between 
$T_{\rm c}^{\rm MCT}$ and $T_{\rm c}^{\rm MD}$ of Fig. \ref{fig:Tc}.
Recalling the three main approximations of MCT, we suggest that the convolution
approximation, Eq. (\ref{eq:convol}), might break for stiff chains. Though possibly it is not the case
for intermolecular contributions, its breakdown for {\it intramolecular} contributions
in stiff chains is plausible.  It is known that the convolution approximation
fails when static correlations show a strong directionality at near-neighbor distances,
as for e.g., network-forming liquids as silica \cite{silicac3}. This directionality is clearly enhanced
for intrachain correlations by increasing the barrier strength, as evidenced by
the progressively larger values of the end-to-end radius (see Table \ref{tab:paramkbkt}). 
For the case of silica, it has been
shown that the explicit inclusion of three-point static correlations in the MCT vertex
improves significantly the quality of the comparison between theory and simulations \cite{silicac3}.
A similar improvement might be achieved in the present case by similarly incorporating
the intrachain three-point static contributions. Work in this direction is in progress.

\begin{center}
{\bf IX. CONCLUSIONS}
\end{center}

We have performed simulations on a simple bead-spring model 
for polymer melts with intramolecular barriers. 
The role of such barriers on the glass transition has been investigated
by systematically tuning the barrier strength. Dynamic correlators probing the structural relaxation
have been analyzed in the framework of the Mode Coupling Theory. 
We have obtained critical nonergodicity parameters, critical
temperatures and dynamic exponents of the theory from consistent 
fits of simulation data to MCT asymptotic laws. 
From the analysis of the critical nonergodicity parameters we deduce that the 
presence of the barriers induces a weaker localization length in the system at fixed density. 
The increase of the barrier strength at fixed density also induces
a higher critical temperature $T_{\rm c}$. The values of the dynamic exponents, 
as obtained from the phenomenological analysis of the simulation data, 
exhibit significant differences between the limit of
fully-flexible and stiff chains. In particular the so-called $\lambda$-exponent
takes standard values $\lambda \sim 0.7 $ for the fully-flexible case 
and values approaching the upper limit $\lambda =1 $ for strong intramolecular barriers. 
While the former $\lambda$-values are characteristic of simple systems
dominated by packing effects, transitions with $\lambda \approx 1$  arise in systems with
different competing mechanisms for dynamic arrest. 
In our systems these large $\lambda$-values suggest a competition
between two distinct mechanisms: general packing effects and polymer-specific intramolecular barriers. 

For a comparison between simulation and theory, we have numerically solved the MCT equations, 
following a recent extension of the MCT by Chong and co-workers 
for polymer melts. The approximations assumed by the structural PRISM theory, 
which are introduced in the MCT equations, are fulfilled for all
the investigated values of the barrier strength.
We have compared  the critical  nonergodicity parameters and critical temperatures $T_{\rm c}$,
as obtained by solving the MCT equations, with the corresponding values from
the phenomenological analysis of the simulation data. 
The theoretical calculations qualitatively reproduce
the trends observed in the simulations for low and moderate barriers.
However strong discrepancies are observed 
as the limit of high barriers is approached. The reason for such
a disagreement possibly lies in the
nature of the approximations made in the derivation of the MCT equations. 
In particular, the convolution approximation for  three-point static correlations might be
unadequate for stiff chains. We suggest that a reformulation of MCT equations 
for polymer melts, explicitly including intrachain three-point static correlations, might lead to a
better agreement between simulations and theory. Work in this direction is in progress.

\begin{center}
{\bf ACKNOWLEDGEMENTS}
\end{center}

We acknowledge financial support from the projects NMP3-CT-2004-502235 (SoftComp, EU), 
MAT2007-63681 (Spain), 2007-60I021 (Spain), and IT-436-07 (GV, Spain).
We thank S.-H. Chong, E. Zaccarelli, M. Fuchs and M. Sperl for useful discussions.


\begin{thebibliography}{99}
\bibitem{hopping} In reality, it is well-known that hopping events, not accounted for within
the ideal version of MCT here described, restore ergodicity and the MCT transition does not really
occur. Indeed $T_{\rm c}$ is found to be higher than the actual `laboratory' glass transition
of the system. Having said this, the MCT asympotic laws (see main text)
for dynamic correlators and relaxation times by approaching $T_{\rm c}$ usually hold
for not too small values of the separation parameter $|T/T_{\rm c} -1|$.
\bibitem{notegridpicard}
Fast Fourier tranforms are used  to switch back and forth between the $r$- and $q$-space.
We use a grid of $G_r=6000$ equispaced points from $r=0$ to $r_{\rm max}=40$ in the real space. 
For the reciprocal space we use $G_q=800$ points from $q=0$ to  $q_{\rm max}=65$.
We use the criterion for convergence 
$G_r^{-1} \sum_{i=1}^{G_r}\{[\Gamma_{j+1}(r_i) -\Gamma_j (r_i) ] ^2 \}^{1/2} < 10^{-7}$.
The iteration is performed by mixing the old solution $\Gamma_j$ 
with the current $\Gamma$ in order to obtain  the new solution $\Gamma_{j+1}$, according to
$\Gamma_{j+1}= 0.2\Gamma   +0.8\Gamma_{j}$.
\end{thebibliography}
\end{document}